\documentclass[accepted=2022-08-03, onecolumn, letterpaper, 11pt]{quantumarticle}
\pdfoutput=1

\usepackage{float}
\usepackage{cite}
\usepackage{amsmath}
\usepackage{braket}
\usepackage[utf8]{inputenc} 
\usepackage[T1]{fontenc}    
\usepackage[pagebackref]{hyperref}
\hypersetup{
    pdftitle={}, 
    pdfauthor={},
    colorlinks=true,
    linkcolor=quantumviolet,
    citecolor=quantumviolet,
    filecolor=quantumviolet,
    urlcolor=quantumviolet
}
\renewcommand{\backref}[1]{}

\renewcommand{\backrefalt}[4]{%
\ifcase #1 %
\or
[p.\ #2]%
\else
[pp.\ #2]%
\fi}

\usepackage{url}            
\usepackage{booktabs}       
\usepackage{amsfonts}       
\usepackage{nicefrac}       
\usepackage{microtype}      
\usepackage{lipsum}
\usepackage{enumitem}
\usepackage{appendix}
\usepackage{graphicx}
\usepackage{mathtools}
\usepackage{subfig}
\usepackage{mathtools}  
\usepackage{pdflscape}
\usepackage{tensor}

\newcommand{\angstrom}{\textup{\AA}}

\usepackage{xcolor}

\def\doi#1{\href{https://doi.org/#1}{doi:#1}}

\usepackage{amsthm}
\newtheorem{definition}{Definition}

\title{Connecting geometry and performance of two-qubit parameterized quantum circuits}

\author{Amara Katabarwa}
\affiliation{Zapata Computing,~Inc., 100 Federal Street, 20th Floor, Boston, Massachusetts 02110, USA}
\email{amara@zapatacomputing.com}
\thanks{These authors contributed equally to this work.}
\orcid{0000-0002-0078-5202}

\author{Sukin Sim}
\affiliation{Harvard University}
\affiliation{Zapata Computing,~Inc., 100 Federal Street, 20th Floor, Boston, Massachusetts 02110, USA}
\email{sukin.sim@zapatacomputing.com}
\thanks{These authors contributed equally to this work.}

\author{Dax Enshan Koh}
\affiliation{Institute of High Performance Computing, Agency for Science, Technology and Research (A*STAR), 1 Fusionopolis Way, \#16-16 Connexis, Singapore 138632, Singapore}
\email{dax\textunderscore koh@ihpc.a-star.edu.sg}
\orcid{0000-0002-8968-591X}

\author{Pierre-Luc Dallaire-Demers}
\affiliation{Zapata Computing,~Inc., 100 Federal Street, 20th Floor, Boston, Massachusetts 02110, USA}
\email{pierre-luc@zapatacomputing.com}

\def\christoffel#1#2{\Gamma^{#1}_{\hphantom{#1}#2}}

\begin{document}
\maketitle

\begin{abstract}
Parameterized quantum circuits (PQCs) are a central component of many variational quantum algorithms, yet there is a lack of understanding of how their parameterization impacts algorithm performance. We initiate this discussion by using principal bundles to geometrically characterize two-qubit PQCs. On the base manifold, we use the Mannoury-Fubini-Study metric to find a simple equation relating the Ricci scalar (geometry) and concurrence (entanglement). By calculating the Ricci scalar during a variational quantum eigensolver (VQE) optimization process, this offers us a new perspective to how and why Quantum Natural Gradient outperforms the standard gradient descent. We argue that the key to the Quantum Natural Gradient's superior performance is its ability to find regions of high negative curvature early in the optimization process. These regions of high negative curvature appear to be important in accelerating the optimization process.
\end{abstract}

\section{Introduction}
\label{sec:introduction}

Parameterized quantum circuits (PQCs) are central components of variational quantum algorithms, a class of algorithms well-suited for being implemented on near-term quantum devices.
In these algorithms, parameters of PQCs are tuned or optimized, using a classical computer, to prepare quantum states on a quantum computer that encode solutions of a problem, e.g.~the ground state of a quantum system or a target probability distribution \cite{cerezo2021variational,bharti2022noisy}.
While PQCs can be constructed leveraging physical insights, 
e.g.~unitary coupled-cluster \cite{Yung2015,Cao2019Quantum,Anand2021Quantum},
in applications such as quantum machine learning or implementation of proof-of-principle experimental demonstrations of variational quantum algorithms, PQCs follow a more heuristic design. 
These PQCs comprise repeated layers of a particular low-depth configuration of single-qubit and two-qubit gate operations \cite{Havlicek2018Supervised, Kandala2017}.
Despite the rapid developments in variational quantum algorithms, PQCs are not yet well understood nor effectively designed.
For instance, while ``hardware-efficient'' circuits may correspond to very low depths, it was shown that many of the parameters (and their corresponding gates) are often unnecessary or redundant \cite{Rasmussen2020Reducing,Sim2020Adaptive,Funcke2021Dimensional}.
In our work, to develop a more concrete understanding of the role of parameters in PQCs, we begin with a mathematical description of PQCs.

Formally, parameterized quantum circuits are maps $\Psi(\boldsymbol\theta)$ between a set of continuous parameters $\boldsymbol\theta$ and the output statistics of a set of observables on a given system.
\begin{figure}[ht]
\centering
\includegraphics[scale=0.33]{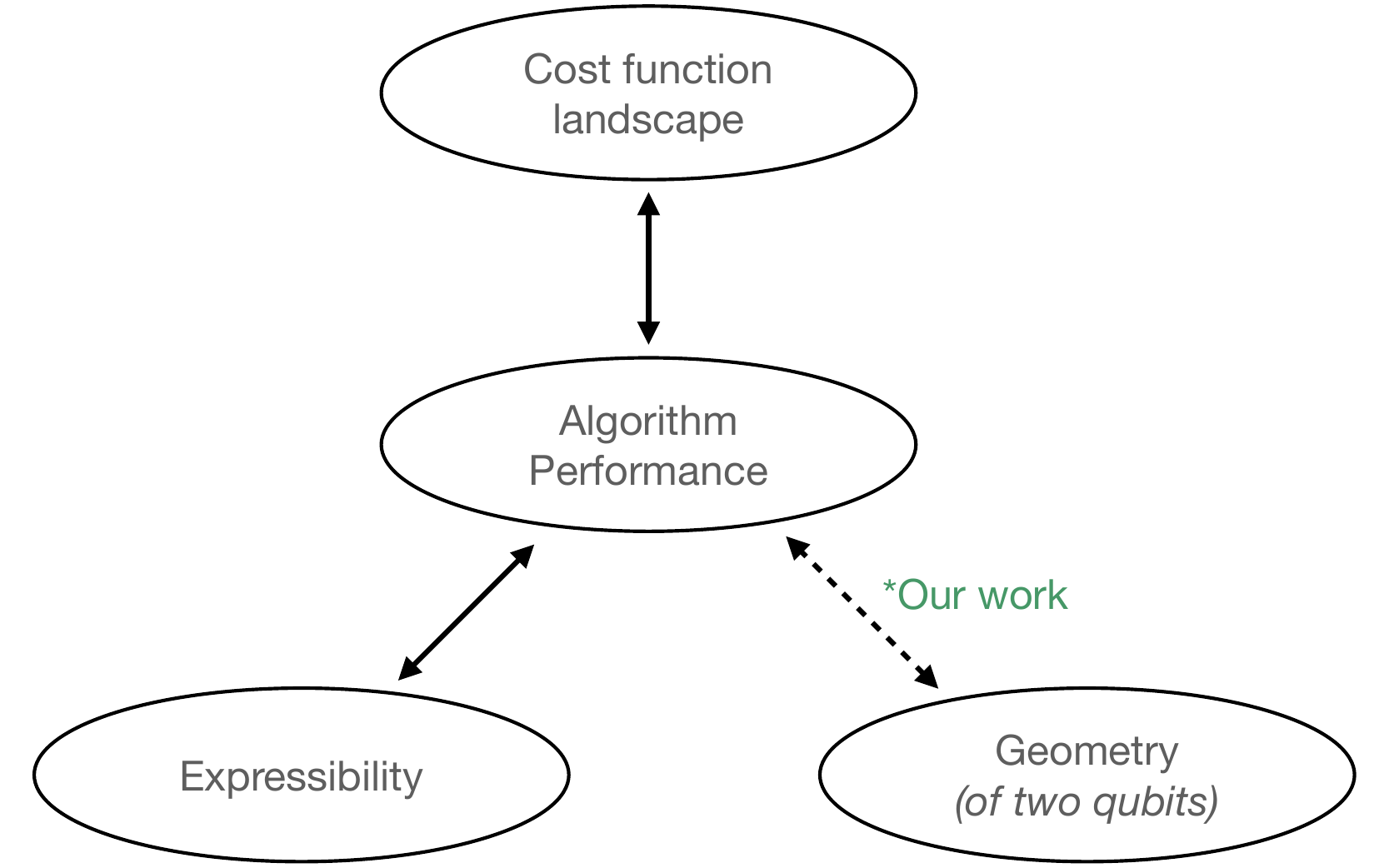}
\caption{Progress on the study of parameterized quantum circuits (PQCs) and their connections to the algorithm performance. Past works have investigated how use of particular circuits leads to features in the cost function landscape that hinder algorithm performance (e.g.~barren plateaus). Other works have introduced circuit metrics such as expressibility and started correlating them to algorithm performance. Our work, while limited to two qubits, initiates the discussion to connect the geometry of PQCs to the algorithm performance and provides valuable insights into why and how Quantum Natural Gradient accelerates the optimization step of near-term quantum algorithms.
}
\label{fig:intro}
\end{figure}
In recent years, significant progress has been made to better understand PQCs and their connection to the algorithm performance as illustrated in Fig.~\ref{fig:intro}.
For instance, Ref.~\cite{mcclean2018barren} introduced the phenomenon of ``barren plateaus'' or regions of non-informative gradients in cost function landscapes resulting from employing PQCs that form approximate 2-designs. Several works extended this work, connecting particular circuit structures to cost function landscape features such as barren plateaus and narrow gorges \cite{arrasmith2021equivalence} that hinder algorithm performance.
For a better understanding and evaluation of PQCs, Ref.~\cite{Sim2019} introduced ``expressibility'' as a quantity to compare among PQCs and rule out circuits with limited capabilities. Since then, works such as Ref.~\cite{hubregtsen2021evaluation} have started correlating expressibility to performance metrics of particular variational quantum algorithms. Additionally, Ref.~\cite{holmes2022connecting} connected high expressibility to the presence of barren plateaus in cost function landscapes.

Among past works connecting PQCs to algorithm performance, Ref.~\cite{Stokes2019}
introduced and investigated the use of Quantum Natural Gradient (QNG) descent to accelerate the optimization of parameters making connections to imaginary time evolution. This modification of the gradient descent which involves an inverse of a metric that is one fourth of the Quantum Fisher Information. The Fubini-Study metric tensor employed in QNG does not contain information about the objective function (unlike gradients or Hessian) and yet has been shown to significantly accelerate optimization by accounting for the geometry of the wave function. Since its discovery, the QNG has been a subject of additional study. For example, while it might suffer from the same \textit{barren plateau}  \cite{mcclean2018barren} problem, there is numerical evidence that at least for shallow circuits the variance of the cost function is orders of magnitude bigger than if vanilla gradient descent \cite{haug2021capacity} was used. In \cite{haug2021optimal}, methods for using the QNG and adaptively choosing learning rates that can derived from the QNG were shown to outperform traditional methods like ADAM and LBFGS in learning quantum states. These performance gains come with an overhead cost of calculating what the QNG is, but work has been done showing that for quantum simulation there exist efficient algorithms to calculate it \cite{jones2020efficient} and even in the setting where one is required to calculate it from measurements coming from quantum hardware the overall cost is asymptotically negligible in the number of iterations and qubits \cite{VanStraaten2021}. The QNG has also been extended to the non-unitary case where some depolarizing noise is allowed in the circuit \cite{koczor2019quantum}.
To gain a deeper understanding of the connection between the geometry of the wave function and algorithm performance, 
we initiate the discussion by providing a geometric characterization of two-qubit PQCs.

The structure of the paper is as follows:
\begin{enumerate}[label= (\roman*) ]
    \item In Sec.~\ref{sec:geometry_intro}, we introduce the geometrical formalism we shall use to analyze PQCs. We consider four specific two-qubit PQCs and take some time to carefully geometrically characterize the circuits. 
    \item In Sec.~\ref{sec:concurrence_curvature}, we introduce the notions of concurrence and the scalar curvature and uncover a simple and remarkable relationship between the two concepts.
    \item In Sec.~\ref{sec:numerical_experiments}, we capitalize on the relationship between concurrence and curvature in order to provide geometrical insights to the VQE optimization process in the PQCs considered in this work. 
    \item In Sec.~\ref{sec:conclusion}, we summarize our work, providing future directions of research and open questions.
\end{enumerate}

\section{Geometry, quantum mechanics, and parameterized quantum circuits}\label{sec:geometry_intro}
A geometric reformulation of quantum mechanics has largely been a mathematical and theoretical pursuit while the algebraic view of quantum mechanics appears to be more practical for the everyday quantum mechanician. 
The question that may arise is whether the geometric  approach remains a mere mathematical happenstance or can be leveraged in more practical day-to-day settings.
For completeness sake, we provide a quick introduction to the geometric structure of quantum mechanics and hopefully lay bare the complications that arise for the projective nature of quantum states.
The discussion will hopefully explain more clearly the geometric connection and what goes wrong if it is not fully incorporated in one's understanding of parameterized quantum circuits.
The K{\"a}hler structure of quantum mechanics furnishes us with a Riemannian metric and symplectic structure encompassed in what is called the \textit{Quantum Geometric Tensor} (for more in-depth pedagogical introductions to this large and rich area, we refer the readers to \cite{Heydari2016,robertgeroch2013}).

\subsection{Quick introduction to fiber bundles}
Most topological spaces have a complicated geometry for which there most likely is no easy intuitive picture. 
One way of addressing this problem is instead to think about the local geometric structure of this complicated topological space.
The local geometry will be thought about as being simply a Cartesian product of lower dimensional geometries.
For the rest of the paper, we are going to assume that the complicated topological space $\mathcal{E}$ can be endowed with a manifold structure.
The idea of a \textit{fiber bundle} is to imagine the geometry of $\mathcal{E}$ as locally being a Cartesian product of two lower dimensional manifolds $\mathcal{B}$ and $\mathcal{F}$.
If in fact the \textit{total space}  $\mathcal{E}$ can be thought of as merely $\mathcal{B} \times \mathcal{F}$ then the fiber bundle is called a \textit{trivial bundle}. Otherwise, in general, the local Cartesian structure will be patched together with some twisting that will make the global geometry look very different from the local geometry. 
The local picture of $\mathcal{E}$ can be thought of as there being a \textit{base space}, $\mathcal{B}$ (which we go to by a projection map $\pi$ from the total space, $\mathcal{E}$) and then at each point on the base space attaching the second geometry called the $\textit{fiber}$, $\mathcal{F}$. We denote the fiber bundle as 

\begin{equation}
    \mathcal{F} \rightarrow  \mathcal{E} \xrightarrow{\pi} \mathcal{B}.
\end{equation}

\begin{figure}[ht!]
    \centering
   \subfloat[Fiber bundle picture of a cylinder]{\includegraphics[width=65mm, height=45mm]{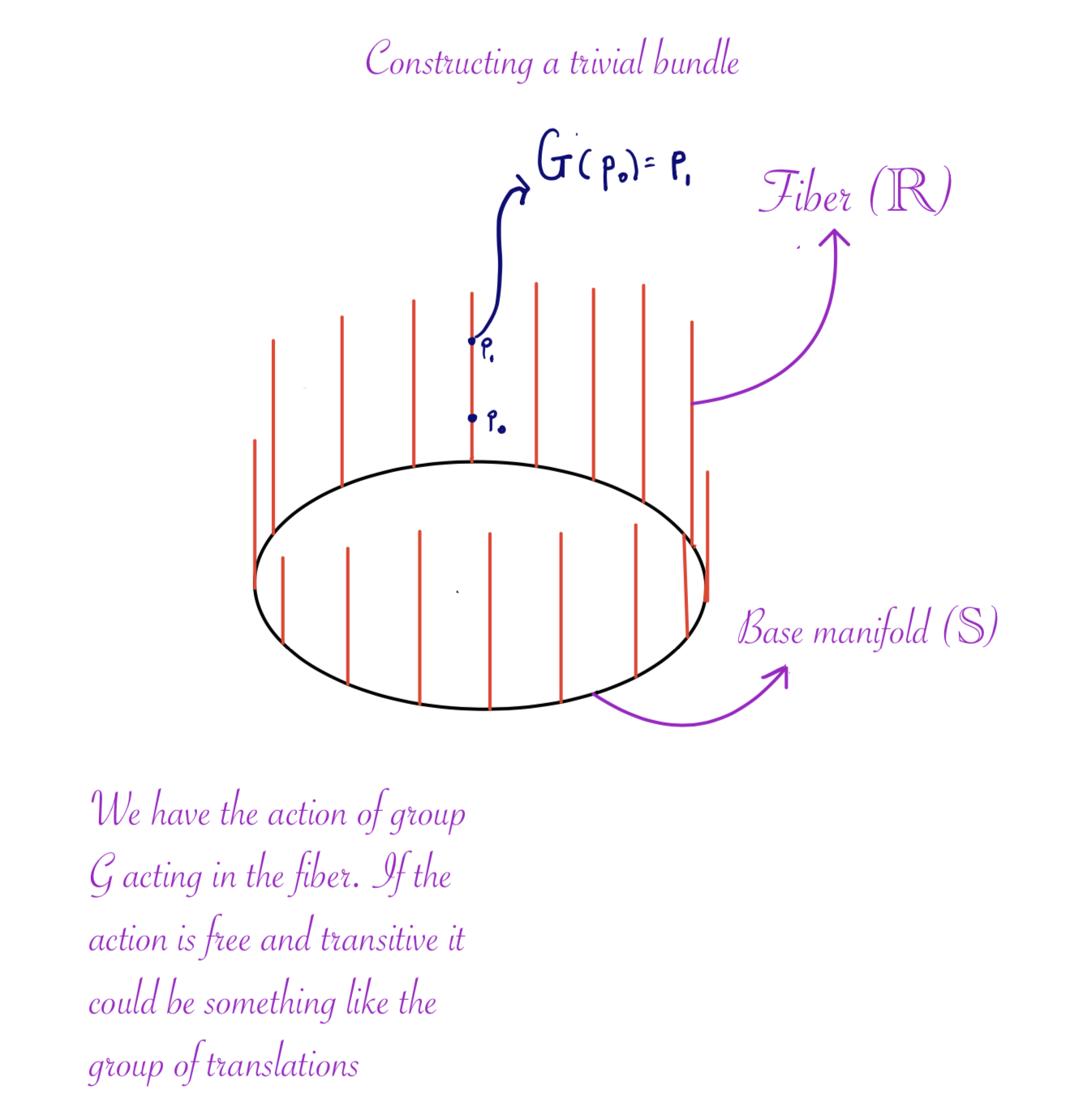}} 
   \hspace{6mm}
   \subfloat[Mobius strip; the lines represent the fibers attached with a global twist]{\includegraphics[width=65mm, height=45mm]{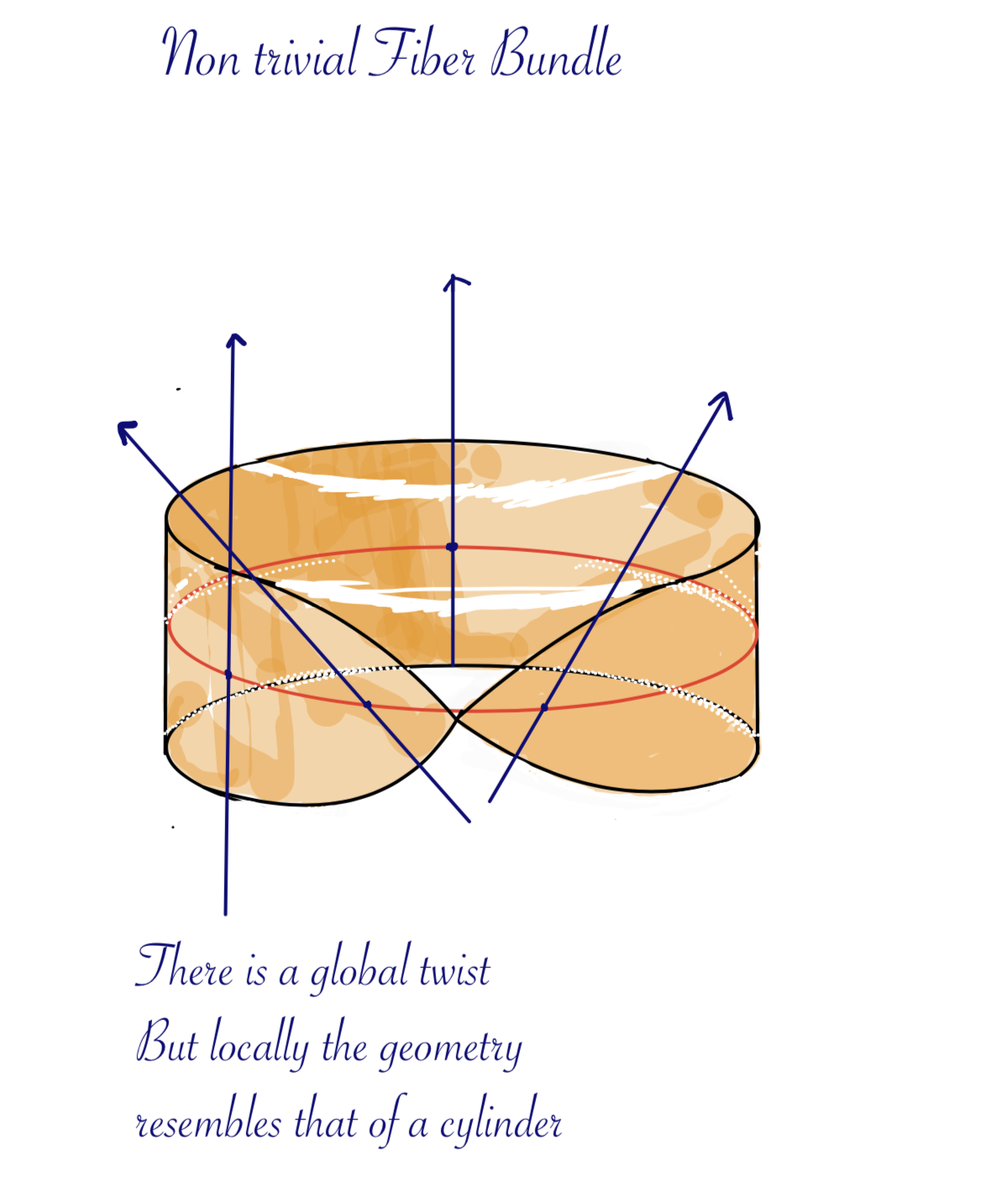}}
   \caption{Fiber bundle perspective of a cylinder (a) and a M\"obius strip (b). The lines in both pictures represent the fibers, while the circle represents the base manifold. The fiber bundle perspective helps us think locally of complicated manifolds in terms of smaller dimensional manifolds.}
   \label{fig:fiber_bundle_cartoon}
\end{figure}

For quantum mechanics, we are interested in fiber bundle for which we have an action of a group $\mathcal{G}$. How does the group act on our fiber bundle? For all $g \in \mathcal{G}$, and for any point $b \in \mathcal{B}$ , $bg$ is merely another point in the fiber attached at the base point $b$. We demand we have continuous right action of the group so that in fact $\mathcal{G}$ is a lie group and that this action is \textit{free} and \textit{transitive} i.e. that respectively only the identity element preserves a point in the fiber, $\mathcal{F}$ and that for any two points, $f, f'$ in the fiber there is a unique group element $g$ such that $f = f'g$. What this amounts to is that the group $\mathcal{G}$ is in fact isomorphic to the fiber $\mathcal{F}$. When this happens the fiber bundle is called a \textit{Principal Fiber Bundle}. This principal fiber bundle is then denoted as 
\begin{equation}
    \mathcal{G} \rightarrow  \mathcal{E} \xrightarrow{\pi} \mathcal{B}.
\end{equation}

\subsubsection{Quantum mechanics and geometry}
\label{sec:qm_geometry}
We are often taught early that global phases do not matter.
While this is usually peppered over since these phases have no physical consequences, it has major consequences from our point of view. 
Consider a wave function in a Hilbert space, $\ket{\psi} \in\mathbb{H}$. The fact that the global phase does not physically matter means that we are instead considering equivalence classes denoted as $[\ket{\psi}]$. 
This means that the physical space is in fact a projectivized Hilbert space $\mathbb{P}(\mathbb{H})$. 
Note that we can also think of this as a principal fiber bundle where the group action is given by $U(1)$ so that the principal fiber bundle we have is 

\begin{equation}
    U(1) \rightarrow  \mathbb{H} \xrightarrow{\pi} \mathbb{P}(\mathbb{H}).
\end{equation}

On the way to introducing a metric on the Hilbert Space $\mathcal{H}$, we recall that we have a Hermitian inner product,  $h :\mathbb{H} \times \mathbb{H} \rightarrow \mathbb{C}$ defined by 
\begin{equation}
\label{hermitianform}
    h(\phi, \psi) = \langle \phi | \psi \rangle  = G(\phi, \psi) + i F(\phi,\psi),
\end{equation}

where $G(\phi, \psi) = \text{Re} (\langle \phi | \psi \rangle)$ and $F(\phi, \psi) = \text{Im} (\langle \phi | \psi \rangle)$. The real part is what produces a Riemannian metric and the imaginary part is what gives us a symplectic structure. To arrive at the metric we shall need to consider the tangent space of $\mathbb{P(H)}$. The tangent space at point on $\mathbb{P(H)}$, $T_{[ \ket{\psi}]}\mathbb{P(\mathbb{H})}$ is the quotient vector space $T_{[ \ket{\psi}]}\mathbb{H}/ $  $ \simeq$ $\mathbb{H}  $ where the equivalence is $\ket{\phi_1} \sim \ket{\phi_2}$ whenever $ \ket{\phi_1} -\ket{\phi_2}   = a \ket{\psi}  $ where $a \in \mathbb{C}$. By choosing an orthogonal complement to the space spanned by $\ket{\psi}$ i.e $\langle \psi | \phi \rangle = 0$, we don't have an ambiguity in the tangent space. To ensure that this condition is always met, we use the projection to the orthogonal complement:
\begin{equation}
    P^{\perp}_{\psi}  = \mathbb{I} - \frac{\ket{\psi}\!\!\bra{\psi}}{\langle \psi | \psi \rangle}.
\end{equation}

On the tangent space remembering $T_{\ket{\psi}}\mathbb{H} \simeq \mathbb{H}$, we see that Eq.~\eqref{hermitianform} may be written in the form
\begin{equation}
\label{hermitianform2}
h(\phi_1, \phi_2)= \bra{\phi_1} P^{\perp}_{\psi}\ket{\phi_2} = \langle \phi_1| \phi_2 \rangle  -  \langle \phi_1| \psi \rangle \langle \psi| \phi_2 \rangle.
\end{equation}

This is in fact what is called the \textit{Quantum Geometric Tensor} in the literature. To see this, consider the wave function $\ket{\psi}$, then we go to the tangent space with bases $\{ \ket{\partial_i \psi}  \} $ and then plugging this into (\ref{hermitianform2}) we get 

\begin{equation}
    h(\partial_i \psi, \partial_j \psi)= \bra{ \partial_i \psi} P^{\perp}_{\psi}\ket{ \partial_j \psi} = \langle  \partial_i \psi | \partial_j \psi \rangle  -  \langle \partial_i \psi | \psi \rangle \langle \psi| \partial_j \psi \rangle.
\end{equation}

\textit{Remark:} Note the intrinsic nature of the above derivation for the Quantum Geometric Tensor. Usually this is arrived at by considering the transition between a state $\ket{\psi_{\vec{u}}}$  and infinitesimally close-by state $\ket{\psi_{\vec{u + \delta u}}}$ then imposing `gauge invariance' of quantum mechanics \cite{cheng2010quantum} or noting that the squared differential, $ds^2 = g_{ij}dx^idx^j$, can in quantum mechanics be written as $ds^2 \propto (1-F^2)$  where $F = | \langle\phi| \psi \rangle|  $\cite{Stokes2019} to lowest order in the parameters that appear in the wave function. The latter derivations, \cite{cheng2010quantum} and \cite{Stokes2019}, while correct obscure its intimate origins to the projective nature of Quantum mechanics. The real part of the Quantum Geometric Tensor in (\ref{hermitianform2}) is the \textit{Fubini-Study metric}. 

\subsection{Fiber bundles and PQCs}
A bit of care must be taken in applying the information geometric view to PQCs. In order to see why, we carefully define what a PQC is. To do this we follow Ref.~\cite{juthohaegeman2014} and define a variational ansatz or PQCs as follows:

\begin{definition}
 A \textbf{\textit{parameterized quantum circuit}} is the image of the following map $\Psi : \mathcal{U} \in \mathbb{R}^m \rightarrow \mathbb{P}(\mathbb{H})$, i.e.
     \begin{equation}
         \mathcal{M} = \Psi(\mathcal{U}):= \{ [\ket{\Psi( \boldsymbol \theta)}] \}
     \end{equation}
 where $ \boldsymbol \theta = (\theta_1, \theta_2, \dots, \theta_m ) \in \mathcal{U} \subset \mathbb{R}^m $.
\end{definition}

To get at the metric we go to the tangent space and as a consequence define the following push-forward map:
\begin{equation}
    d\Psi(\boldsymbol v) : T_p \mathbb{R} \rightarrow T_{[\ket{\Psi(\vec{\theta})}]}\mathcal{M} : v^i \partial_{\theta_i} \longrightarrow v^i \partial_{\theta_i} [\ket{\Psi(\boldsymbol \theta)}]= v^i [\ket{ \partial_{\theta_i}\Psi(\boldsymbol \theta)}]
\end{equation}
where $\boldsymbol v = v^i \partial_{\theta_i}$. This allows us to define the metric on $\mathbb{P}(\mathbb{H})$ using our points in $\mathbb{R}^m$ and this is done by using the pull-back by $\Psi$, back to $T_p\mathbb{R}^m$  i.e. 
\begin{equation}
  \label{metricdef}
    g_{\boldsymbol \theta}: T_p\mathbb{R}^m \times T_p\mathbb{R}^m \rightarrow \mathbb{R} : (\boldsymbol v_1,\boldsymbol v_2) = g(\boldsymbol v_1,\boldsymbol v_2) = g(d\Psi(\boldsymbol v_1), d\Psi(\boldsymbol v_2))
\end{equation}
where we define the value of the metric as 
\begin{equation} 
\label{metricvalue}
    g(\boldsymbol v_1,\boldsymbol v_2)  =   v_1^i v_2^j(  \langle\partial_i \Psi| \partial_j \Psi \rangle - \langle \partial_i \Psi | \Psi \rangle \langle \Psi | \partial_j \Psi \rangle).
\end{equation}
Here, we have suppressed the dependence on $\boldsymbol \theta$ for ease of notation; we have implicitly chosen a point on a ray in evaluating the value for the metric.

Now the map $\Psi$ need not be injective so that in fact the metric in Eq.~\eqref{metricvalue} is, in general, degenerate. This is in fact what happens with the circuits we study in this work. 
As a consequence, no unique inverse metric exists.
The problem derives from the fact that the parameters in the wave function are not chosen with the projective nature of quantum mechanics and thus when we ask physical questions like expectation values of hermitian operators we will in fact live on some constrained surface in $\mathbb{P}(\mathbb{H})$; in other words our tangent space at some point will in general be of smaller dimension than $T_p\mathbb{R}^m $.

Naively, we have the following Quantum Geometric Tensor for our PQC:
\begin{equation}
    \label{ParametrizedQGT}
    G_{\theta_i \theta_j} = \langle\partial_{\theta_i} \Psi(\boldsymbol\theta)|{\partial_{\theta_j} \Psi(\boldsymbol\theta)} \rangle - \langle\partial_{\theta_i} \Psi(\boldsymbol\theta)| \Psi(\boldsymbol\theta) \rangle 
    \langle  \Psi(\boldsymbol\theta)| \partial_{\theta_j} \Psi(\boldsymbol\theta) \rangle.
\end{equation}

\subsubsection{Two-qubit parameterized quantum circuits and geometry}

In this section we study how the PQCs fit in the geometry of $\mathbb{P}(\mathbb{H})$ and find the constrained surfaces they live on. This exercise elucidates what part of the geometry the PQCs have access to.

There will be four major circuits we shall consider in studying the information-geometric view of PQCs, namely:
\begin{enumerate}
    \item the \textit{Hardware-Efficient Ansatz} (HEA) introduced by \cite{Kandala2017} and studied in the context of the Fubini-Study metric by \cite{Yamamoto2019},
    \item the \textit{Low Depth Circuit Ansatz} (LDCA) introduced in \cite{Dallaire-Demers2018},
    \item an ansatz introduced in the context of quantum generative adversarial networks (QGANs) \cite{QGAN_PLDD2018}, and 
    \item an ansatz used in \cite{Dallaire-Demers2020} composed of gates native to the Sycamore chip \cite{Arute2019}, e.g. the fermionic simulation (fSim) operation. We will refer to this ansatz as the Sycamore HEA (sHEA).
\end{enumerate}

\begin{figure}[ht]
\centering
\subfloat[Hardware-efficient ansatz (HEA)]{\includegraphics{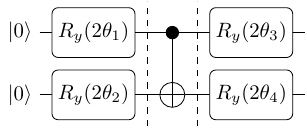}} \hspace{6mm}
\subfloat[QGAN ansatz]{\includegraphics{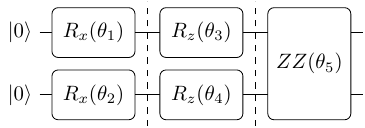}} \\
\vspace{2mm}
\subfloat[Low-depth circuit ansatz (LDCA)]{ \includegraphics{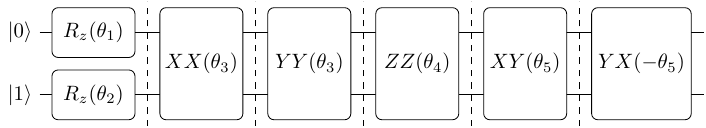}}\\
\vspace{2mm}
\subfloat[Sycamore hardware-efficient ansatz (sHEA)]{ \includegraphics{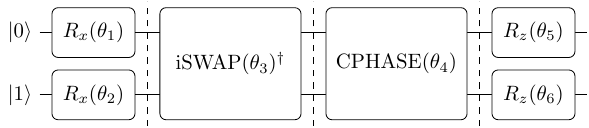}}\\
\caption{Two-qubit circuit blocks considered in this work. Dashed lines indicate different circuit layers or moments. Gate definitions are provided in Appendix~\ref{app:gate_definitions}.}
\label{fig:circuit_diagrams_v2}
\end{figure}

\noindent In general, we have the following isomorphisms
\begin{equation}
 \label{generalisomorphisms}
    \mathbb{P}(\mathbb{H}^{n+1}) \simeq S^{2n-1}/S^1 \simeq \mathbb{C}P^n \simeq G_{1, n+1}
\end{equation}
where $\mathbb{P}(\mathbb{H}^{n+1})$ is the projectivized version of an $n+1$ dimensional Hilbert space, $\mathbb{C}P^n$ is the complex projective space and $G_{1, n+1}$ is the Grassmanian of dimension $n$. For two qubits, the string of isomorphisms in  Eq.~\eqref{generalisomorphisms} become specifically the following isomorphisms:

\begin{equation}
 \label{twoqubitisomorphisms}
    \mathbb{P}(\mathbb{H}^{4}) \simeq S^{7}/S^1 \simeq \mathbb{C}P^3 \simeq G_{1, 4}.
\end{equation}

Luckily, for two and incidentally for three qubits, we have a fiber bundle picture namely the Hopf Fibration. For two qubits, the Hopf Fibration can be thought of as an $SU(2)$ principal fiber bundle. 
We now concentrate on $S^7$ and think of it having a Hopf Fibration. It is a well-known fact that the fibration has the following character:

\begin{equation}
    S^3 \rightarrow S^7 \xrightarrow{\pi} S^4
\end{equation}
The fiber $S^3$ is isomorphic to SU(2). 

From a quantum mechanical point of view $S^4$,which can locally be split into two spheres, i.e $S^2 \times S^2$, can be given the following interpretation: the first sphere is a \textit{quasi}-Bloch sphere representing the degrees of freedom one observer has access to for a given two-qubit entangled state and the second sphere parameterizes the amount of entanglement shared by the two qubits \cite{wie2020bloch}.

For each of the  major circuits, we calculate the version of the metric that is in general degenerate, a metric that is specifically tailored to the geometry of $S^7$ and provide explicit parametrizations of how the two qubit circuits sit inside the geometry.

The Hopf base and fiber parameters are calculated as follows \cite{wie2020bloch, levay2004geometry}.
For a 2-qubit state:
\begin{align}
    \ket\psi = \alpha\ket {00}
    + \beta\ket {01}
    + \gamma\ket {10}
    + \delta\ket {11}
\end{align} where $\alpha,\beta,\gamma,\delta \in \mathbb C$,
compute the following:
\begin{enumerate}
    \item First, calculate the $S^4$ Hopf base parameters $(\theta_A,\phi_A)$ and $(\chi,\xi)$ and the equivalent Cartesian coordinates $(x_0,x_1,x_2,x_3,x_4)$ as follows:
    \begin{enumerate}
    \item $x_0 = |\alpha|^2 +
    |\beta|^2 - |\gamma|^2 - |\delta|^2 
    $
    \item $\theta_A = \arccos(x_0)$.
    \item $x_1 = 2\, \mathrm{Re}(\bar\alpha\gamma+\bar\beta\delta)$
    \item $x_4 = 2\, \mathrm{Im}(\bar\alpha\gamma+\bar\beta\delta)$
    \item $\phi_A = \arccos\left(x_1 \csc(\theta_A)\right)$
    \item $\chi = \arccos(x_4\csc(\theta_A)\csc(\phi_A))$
    \item $x_3=2 \,\mathrm{Re}(\alpha\delta-\beta\gamma)$
    \item $x_2=-2 \,\mathrm{Im}(\alpha\delta-\beta\gamma)$
    \item $\xi = \arctan(x_3/x_2)$
\end{enumerate}

In calculating how the PQCs fit inside the fibers using intrinsic co-ordinates, one needs to solve highly non-linear trigonometric equations. We simplify our task, by switching co-ordinates to the constrained extrinsic co-ordinates. We take advantage of the following parameterization equivalence:

\begin{equation}
    \ket{\psi_q} = \begin{pmatrix}
         \cos\theta_A \\
         \sin \theta_A e^{t \phi_A}
       \end{pmatrix}q = \frac{1}{\sqrt{2}} \begin{pmatrix}
          \sqrt{1\pm \sqrt{1-|z|^2-|w|^2}} \\
          \sqrt{1\mp \sqrt{1-|z|^2-|w|^2}} \frac{z+wj}
    {\sqrt{|z|^2+|w|^2}} 
       \end{pmatrix} q, 
\end{equation}

where $q $ is a unit quaternion used to parameterize the fiber and $e^{t \phi_A}$ is a unit quaternion used to parameterize a point on the entanglement sphere.

\item Next, calculate the $S^3$ Hopf fiber parameters, which are represented by the quaternion $q$.
\begin{enumerate}
    \item $z=\tfrac 12(x_1+ix_4)$
    \item $w=\tfrac 12(x_3-ix_2)$
    \item $\gamma_{\pm} = \sqrt{1\pm \sqrt{1-|z|^2-|w|^2}}$
    \item $\ket{\psi_H} =
    (\alpha+\beta j)\ket 0 +
    (\gamma+\delta j)\ket 1
    $
    \item $\ket {c_\pm} = \frac 1{\sqrt 2} \begin{pmatrix}
    \gamma_{\pm} \\
    \gamma_{\mp} \frac{z+wj}{
    \sqrt{|z|^2+|w|^2}
    }
    \end{pmatrix}
    $
    \item $q_\pm = \langle c_\pm|\psi_H\rangle$
\end{enumerate}
\end{enumerate}

\begin{figure}[ht]
    \centering
 \subfloat[Quasi bloch sphere]{\includegraphics{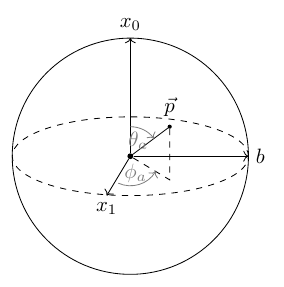}} \hspace{6mm}
 \subfloat[Entanglement sphere]{\includegraphics{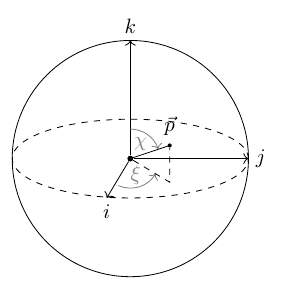}} \\
    \caption{Local geometry of $S^4$.}
    \label{fig:my_label}
\end{figure}

We compute the above Hopf base and fiber parameters for the circuit blocks in Fig.~\ref{fig:circuit_diagrams_v2} and present our results in Appendix
\ref{app:hopfBaseFiber}.

\subsubsection{Local geometric expressibility}
The notion of expressibility for PQCs has been explored in \cite{Sim2019}. By thinking of the Hilbert space from the geometric point of view, namely as $S^7$, we can arrive at a picture of \textit{local geometric expressibility}.  From this perspective we ask ourselves which points in $S^7$ can our PQCs reach, and which constrained surfaces do they live on in terms of some un-constrained co-ordinates. 
As discussed earlier, the local degrees of freedom accessible to an observer are parameterized by one of the spheres with intrinsic coordinates $(\theta_A, \phi_A)$ on the base manifold while the entanglement properties are parameterized by the second sphere parameterized by intrinsic coordinates $(\chi, \xi)$.

{ 
\renewcommand{\arraystretch}{1.5}
\begin{table}[ht]
    \centering
    \caption{PQC base manifold geometry}
    \begin{tabular}{ |p{2.7cm}||p{4cm}|p{4cm}|p{3.2cm}| }
        \hline
        \centering \textbf{Quantum Circuit} & \centering $\boldsymbol{S^2}$ \textbf{(local sphere)} & \centering $\boldsymbol{S^2}$ \textbf{(entanglement sphere)} & \textbf{Constrained Base Space Geometry} \\
        \hline
        \centering$\text{HEA}$ & Explores with $(\phi_A, \theta_A)$ & Does not explore (Stuck to just a point)  & $S^2$ \\
        \hline
        \centering$\text{LDCA}$ & Explores  with $(\theta_A)$ & Explores with $\xi$ & $S^2$ \\
        \hline
        \centering$\text{QGAN}$ & Explores  with $(\phi_A, \theta_A)$ & Explores with $\chi$  & $S^3$   \\
        \hline
        \centering $\text{sHEA}$ & Explores  with $(\phi_A, \theta_A)$ & Explores with $(\chi, \xi)$ & $S^4$  \\
        \hline
    \end{tabular}
    \label{tab:pqc_base_manifold_geometry}
\end{table}
} 

The sHEA circuit covers the largest surface area of points, while we see that both HEA and LDCA cover points on a two sphere. Interestingly although both HEA and LDCA cover points constrained to live on $S^2$, LDCA uses one of its co-ordinates to explore the entanglement sphere. 
This, in principle, allows LDCA to explore more sub-manifolds of different entanglement in $S^4$.

Next we consider how the different circuits differ in the fiber space. Points in the fiber space appear from one observer's point of view as a difference in the gauge i.e they do not change the entanglement properties on the quantum state (This can be made more explicit by introducing a connection). To help abstract out the minutiae we first re-express (\ref{hea_fiber}), (\ref{ldca_fiber}), (\ref{qgans_fiber}) that represent the parameterization in the fiber space in a way that makes physical interpretation easier. For a fixed fiber, these functions will in general depend on a subset of the parameters in the quantum circuit as one moves in that fiber. 
Using the algebraic equivalence between quaternions and the Lie algebra of $SU(2)$ namely:
$$ i \longmapsto i \sigma^x, j \longmapsto i \sigma^y, k \longmapsto i\sigma^z $$

We re-write (\ref{hea_fiber}), (\ref{ldca_fiber}), (\ref{qgans_fiber}) respectively  as follows:
\begin{align}
    q_{\pm HEA} &= \frac{1}{2 \lambda_2} \left(e_{\pm}(\theta_4)I  + i f_{\pm}(\theta_4)\sigma^y \right) \\ 
    q_{\pm LDCA} &= \frac{1}{2 \mu_3} i\left(g_{\pm}(\theta_1, \theta_2, \theta_3, \theta_4, \theta_5)\sigma^y  +  h_{\pm}(\theta_1, \theta_2, \theta_3, \theta_4, \theta_5)\sigma^z \right) \\ 
  \begin{split}
   q_{\pm QGANS} &= \frac 1{\sqrt 2} \left(\sin \left(\frac{\theta _1}{2}\right) \gamma_\mp+\cos \left(\frac{\theta _1}{2}\right) \gamma_\pm \right) i \big( a_{\pm}(\theta_2, \theta_3, \theta_4, \theta_5)I  +  b_{\pm}(\theta_2, \theta_3, \theta_4, \theta_5)\sigma^x \\
   &\quad + c_{\pm}(\theta_2, \theta_3, \theta_4, \theta_5)\sigma^y +  d_{\pm}(\theta_2, \theta_3, \theta_4, \theta_5)\sigma^z \big)
   \end{split} \\
   q_{\pm sHEA} &= i \big( m_{\pm}(\theta_1, \theta_2, \theta_3, \theta_5, \theta_6)I + m_{\pm}(\theta_1, \theta_2, \theta_3, \theta_5, \theta_6)\sigma^x \nonumber\\ 
   &\quad + r_{\pm}(\theta_1, \theta_2, \theta_3, \theta_5, \theta_6)\sigma^y + s_{\pm}(\theta_1, \theta_2, \theta_3, \theta_5, \theta_6)\sigma^z
   \big).
\end{align}
In the equations above only the  $\theta_i 's$ shown are those that can vary in a specific fiber on a point in the base manifold. 
Expressions for $a-g, m, n, s$ can be found in Appendix \ref{app:hopfBaseFiber}.

{ 
\renewcommand{\arraystretch}{2}
\begin{table}[ht]
    \centering
    \caption{PQC Geometry Summary Table
    }
    \begin{tabular}{ |p{2.2cm}||p{12cm}| }
        \hline
        \textbf{Quantum Circuit} & \textbf{Fiber Space}\\
        \hline
        $\text{HEA}$ & $\frac{1}{2 \lambda_2} \left(e_{\pm}(\theta_4)I  + i f_{\pm}(\theta_4)\sigma^y \right)$ \\
        \hline
        $\text{LDCA}$ & $\frac{1}{2 \mu_3} i\left(g_{\pm}(\theta_1, \theta_2, \theta_3, \theta_4, \theta_5)\sigma^y  +  h_{\pm}(\theta_1, \theta_2, \theta_3, \theta_4, \theta_5)\sigma^z \right)$ \\
        \hline
        $\text{QGAN}$ & $ \begin{multlined} \frac 1{\sqrt 2} \left(\sin \left(\frac{\theta _1}{2}\right) \gamma_\mp+\cos \left(\frac{\theta _1}{2}\right) \gamma_\pm \right) i(a_{\pm}(\theta_2, \theta_3, \theta_4, \theta_5)I  \\ +  b_{\pm}(\theta_2, \theta_3, \theta_4, \theta_5)\sigma^x  +  
    c_{\pm}(\theta_2, \theta_3, \theta_4, \theta_5)\sigma^y +  d_{\pm}(\theta_2, \theta_3, \theta_4, \theta_5)\sigma^z  ) \end{multlined}$ \\
    \hline
        $\text{sHEA}$ & $ \begin{multlined} i \big( m_{\pm}(\theta_1, \theta_2, \theta_3, \theta_5, \theta_6)I + m_{\pm}(\theta_1, \theta_2, \theta_3, \theta_5, \theta_6)\sigma^x + \\ 
   r_{\pm}(\theta_1, \theta_2, \theta_3, \theta_5, \theta_6)\sigma^y + s_{\pm}(\theta_1, \theta_2, \theta_3, \theta_5, \theta_6)\sigma^z
   \big) \end{multlined}$  \\
    \hline
    \end{tabular}
    \label{tab:pqc_fiber_space}
\end{table}
} 

\subsection{Quantum Natural Gradient descent}
\label{sec:qng_descent}

We have discussed the possibility of the metric derived for directly applying Eq.~\eqref{ParametrizedQGT} being generally degenerate.  
In this section, we calculate the ``Fubini-Study'' metrics (by following Eq.~\eqref{ParametrizedQGT}) and see that most are indeed degenerate. 
Although this is the case, it has been shown in \cite{Stokes2019} that these degenerate matrices, in practice, lead to improvements in the number of iterations for the optimization process, through approximating to a block-diagonal or diagonal form, adding numerically small values to make matrices invertible, and/or taking the pseudo-inverse.
A more recent study, however, notes that (block-)diagonal approximations to the metric tensor may not be necessary 
as the cost of computing the metric tensor is asymptotically negligible \cite{VanStraaten2021}.
Using the standard gradient descent method, the parameter update rule for the parameters in the quantum circuit is:
\begin{equation} 
\label{gradient_descent}
    \boldsymbol\theta^{(t+1)} = \boldsymbol\theta^{(t)} - \eta \frac{\partial \mathcal{L}(\boldsymbol\theta)}{\partial \boldsymbol\theta}.
\end{equation}
where $\eta$ is the step size or learning rate and $\mathcal{L}(\boldsymbol\theta)$ is the objective function to be minimized.

For the four circuit types considered, we have the following Fubini-Study metrics assuming (\ref{ParametrizedQGT}) and naively calculating them.
Matrix elements that are included in the block-diagonal approximation to the metric tensor are in {\color{blue}blue}.

\begin{align}
  g^{\mathrm{HEA}} &=   \begin{pmatrix}
           \color{blue}{1} & \color{blue}{0} & \langle \sigma^x_2 \rangle_1 & 0 \\
           \color{blue}{0} & \color{blue}{1} & 0  & \langle \sigma^z_1 \rangle _1 \\
           \langle \sigma^x_2 \rangle_1 & 0 & \color{blue}{1} & \color{blue}{\frac{ \langle \{ \sigma^y_{1},\sigma^y_{2}  \} \rangle_2}{2}} \\
           0&  \langle \sigma^z_{1} \rangle_1 & \color{blue}{\frac{\langle \{ \sigma^y_{(1)},\sigma^y_{(2)}  \} \rangle_2}{2}}  & \color{blue}{1} 
    \end{pmatrix} \label{eq:fs_metric_tensor_hea} \\
    g^{\mathrm{LDCA}} &=   \begin{pmatrix}
           \color{blue}{0} & \color{blue}{0} & 0& 0 & 0  \\
           \color{blue}{0} & \color{blue}{0} & 0  & 0 & 0 \\
           0 & 0 & \color{blue}{4} & 0 & 0 \\
              0 & 0 & 0  & \color{blue}{0} & 0 \\
           0&0 &0&0& \color{blue}{1 + \frac{\langle \{\sigma^z_1, \sigma^z_2\}\rangle_4}{2}}
    \end{pmatrix} \label{eq:fs_metric_tensor_ldca}\\
      g^{\mathrm{QGAN}} &=   \begin{pmatrix}
           \color{blue}{1} & \color{blue}{0} & 0 & 0 & 0  \\
           \color{blue}{0} & \color{blue}{1} & 0  & 0 & 0 \\
           0 & 0 & \color{blue}{1- \langle \sigma^z_1\rangle_1^2} & \color{blue}{0} & \langle \sigma^z_2\rangle_2 -\langle \sigma^z_1\rangle_1 \langle \sigma^z_1\sigma^z_2\rangle_1 \\
              0 & 0 & \color{blue}{0}  & \color{blue}{1- \langle\sigma^z_2\rangle^2_1} & \langle \sigma^z_1\rangle_2 -\langle\sigma^z_2\rangle_1\langle\sigma^z_1\sigma^z_2\rangle_1 \\
           0 & 0 & \langle \sigma^z_2\rangle_2 -\langle\sigma^z_1\rangle_1 \langle \sigma^z_1\sigma^z_2\rangle_1  &\langle\sigma^z_1\rangle_2 -\langle \sigma^z_2\rangle_1 \langle\sigma^z_1 \sigma^z_2\rangle_1  & \color{blue}{1- \langle \sigma^z_1 \sigma^z_2 \rangle_1^2}
    \end{pmatrix} \label{eq:fs_metric_tensor_qgan} \\
    g^{\mathrm{sHEA}} &=  \begin{pmatrix}
    \color{blue}{1} & \color{blue}{0} & 0 & 0 & A & -A\\
    \color{blue}{0} & \color{blue}{1} & 0 & 0 & B &  -B  \\
    0 & 0 & \color{blue}{C} & D & E & -E \\
    0 & 0 & D & \color{blue}{F} & -G & G \\
    A  & B & E & -G & \color{blue}{H} & \color{blue}{I} \\
    -A & -B & -E & G & \color{blue}{I} & \color{blue}{H} \\
    \end{pmatrix} \label{eq:fs_metric_tensor_shea} \\
\end{align}

where for $g^{sHEA}$ we have
\begin{align*}
    A &= \ ^1\!\langle \sigma^x_1 \mathcal{S}(\theta_3)^{\dagger} \sigma^z_1 \rangle_2 - \langle \sigma_1^x \rangle_0 \langle \sigma^z_1 \rangle_2 \\
    B &=\ ^1\!\langle \sigma^x_2 \mathcal{S}(\theta_3)^{\dagger} \sigma^z_1 \rangle_2 - \langle \sigma_2^x \rangle_0 \langle \sigma^z_1 \rangle_2  \\
    C &= \frac{1}{2}\left( 1 - \langle \sigma^z_1 \sigma^z_2 \rangle \right) - \frac{1}{4}  \left( \langle \sigma^x_1 \sigma^x_2 + \sigma^y_1 \sigma^y_2 \rangle_1 \right)^2 \\
    D &= \frac{1}{8} \left( -\langle (\sigma^x_1 \sigma^x_2 + \sigma^y_1\sigma^y_2)(I_1 - \sigma^z_1)(I_2- \sigma^z_2) \rangle_2 + \langle\sigma^x_1 \sigma^x_2 + \sigma^y_1 \sigma^y_2 \rangle_1 \langle (I_1 - \sigma^z_1)(I_2- \sigma^z_2 \rangle_2 \right) \\
    E &= \frac{1}{2} \left( \langle ( \sigma^x_1 \sigma^x_2 + \sigma^y_1 \sigma^y_2) \sigma^z_1 \rangle_2 + \langle\sigma^x_1 \sigma^x_2 + \sigma^y_1 \sigma^y_2 \rangle_1 \langle \sigma^z_1 \rangle_2 \right) \\
    F &= \frac{1}{4}\langle (I_1 - \sigma^z_1)(I_2- \sigma^z_2 \rangle_2 - \frac{1}{16}\langle (I_1 - \sigma^z_1)(I_2- \sigma^z_2 \rangle_2^2 \\
    G &= \frac{1}{4}\langle (I_1 - \sigma^z_1)(I_2- \sigma^z_2 \rangle_2 - \frac{1}{4}\langle (I_1 - \sigma^z_1)(I_2- \sigma^z_2 \rangle_2 \langle \sigma^z_1 \rangle_2 \\
    H &= 1 - \langle \sigma^z_2 \rangle_2^2
    I = \langle \sigma^z_1 \sigma^z_2 \rangle_2 - \langle \sigma^z_1 \rangle_2 \langle \sigma^z_2 \rangle_2. 
\end{align*}

\textbf{Note:} The notation used above $\langle \mathcal{O} \rangle_i $ is used to mean the expectation value of the operator $ \mathcal{O}$ with respect to the wave function at the $i^{th}$ layer of the circuit, while $ ^j \!\langle \mathcal{O} \rangle_i = \bra{\psi_j} \mathcal{O} \ket{\psi_i} $ and $\ket{\psi_i}$ is the wave function at the $i^{\textup{th}}$ layer.

Of the four metrics, ones for the QGAN circuit and sHEA are non-degenerate. 
From the previous section, we can see why this is true; we have 3 independent parameters in the base space, 3 parameters in the fiber space but we also have a global constraint that all the amplitudes must add to 1 so that we have a total of 5 independent global parameters. 
This matches the number of parameters in the quantum circuit. The same can be seen with the sHEA geometry where we get an extra degree of independence from the base space.

In practice, $g$ may be singular due to redundant parameterization, or inverting $g$ becomes computationally challenging with increasing number of parameters. 
Thus, in the case of natural gradient descent for training classical neural networks with many parameters, the inverse of the Fisher Information Matrix (FIM) has been approximated as an inverse of a block-diagonal matrix (and further approximated as Kronecker products of inverses of smaller matrices) to reduce the computational overhead \cite{martens2015optimizing}.
The FIM of deep linear networks are also often singular due to parameter redundancies. Thus, generalized inverses are used.
Ref.~\cite{bernacchia2018exact} showed that for deep linear networks, any choice of generalized inverse was effective in accelerating natural gradient descent.

\section{Concurrence and the Ricci Scalar}\label{sec:concurrence_curvature}
Unlike its classical analogue, the Quantum Natural Gradient has no obvious connection to the geometry encountered in the optimization process.
One might have considered calculating the curvature from the Fubini-Study metric derived from the real part of the Quantum Geometric Tensor, but as has been previously argued, one does not get a legitimate metric since the metric used in Quantum Natural Gradient is in general degenerate.
As a consequence, one could consider regularizing the metric either by considering the pseudo-inverse or by adding a small constant multiplied by the identity matrix ($\epsilon I$), a kind of Tikhonov regularization. The downside to this path is that the value of the curvature depends on the regularization procedure and in fact for the Tikhonov regularization  the curvature in some cases is ill-defined since the value at a point depends on how this small constant is brought to zero. Nevertheless, in practice, carefully setting regularization has been shown to reduce the measurement costs of QNG \cite{VanStraaten2021}.

What we opt for therefore, is a rather indirect way of seeing the Quantum Natural Gradient geometrically at work. We will calculate another metric, a \textit{quaternionic} Fubini-Study metric that can be placed on the base manifold of two qubits, $S^4$. 
Amazingly enough this metric connects the concurrence (entanglement) of the quantum circuit and the curvature of the base manifold. 

\subsection{Concurrence}
In the case of two qubits, the entanglement entropy is the unique measure of entanglement.
In work by Hill and Wootters \cite{Scot_Hill1997}, it was noted that the entanglement entropy can be thought of as a function of the concurrence which is defined in the following manner:

\begin{equation}
    C(\psi) = 2|\alpha \delta- \beta \gamma|,
\end{equation}
where $\ket{\psi}=\alpha\ket{00}+\beta\ket{01}+\gamma\ket{10}+\delta\ket{11}$.
The concurrence has been measured in experiments \cite{lichenmingyanglihuazhangzhuoliangcao2017,Zhou2015ConcurrenceMF}. 
Extensions to mixed density matrices can be considered by calculating the convex roof extension, but in this work we stick to pure states when calculating the measure. 

We can also formulate the notion of concurrence within the geometric picture thus far outlined. First note that the concurrence can be written in terms of the extrinsic co-ordinates of  $S^4$ as:
\begin{equation}
    \label{concurrence_geometry}
    C(\psi) = \sqrt{|x_2|^2 + |x_3|^2} .
\end{equation}

The concurrences for the four circuits in terms of their corresponding circuit parameters are:

\begin{align}
    C_{\text{HEA}} &= |\text{sin}(2\theta_1)\text{cos}(2\theta_2)|, \label{eq:hea_concurrence}\\
    C_{\text{LDCA}} &= \frac{1}{2} \sqrt{3 - 2\text{cos}(4\theta_3)\text{cos}^2(2\theta_5) - \text{cos}(4\theta_5)}, \label{eq:ldca_concurrence}\\
    C_{\text{QGAN}} &=  |\text{sin}(\theta_1)\text{sin}(\theta_2)\text{sin}(\theta_5)|, \label{eq:qgan_concurrence}\\
    C_{\text{sHEA}} &= \frac{1}{2} \biggl\{\sin ^2\left(\theta _1\right) \sin ^2\left(\theta _2\right) \left(\cos \left(\theta _3\right)-\cos \left(\frac{\theta _4}{4}\right)\right){}^2 \\
    &\qquad +\left(\sin \left(\theta _3\right)-\sin \left(\theta _1\right) \sin \left(\theta _2\right) \sin \left(\frac{\theta _4}{4}\right)+\sin \left(\theta _3\right) \cos \left(\theta _1\right) \cos \left(\theta _2\right)\right){}^2\biggr\}^{\!1/2}. 
    \label{eq:shea_concurrence}
\end{align}

\subsection{Ricci Scalar}
\label{sec:ricci_scalar}

In Riemannian geometry, the scalar curvature, also known as the \textit{Ricci scalar}, is an invariant that characterizes the curvature of a Riemannian manifold. It may be defined in terms of the metric tensor $g$ as follows: Let $g_{ab}$ denote the components of $g$, and let $g^{ab}$ denote the components of its inverse $g^{-1}$. The \textit{scalar curvature} is defined as
\begin{align}
    \mathcal R = g^{ab}\left(\christoffel c{ab,c} - \christoffel c{ac,b} + \christoffel d{ab}\christoffel c{cd} - \christoffel d{ac}\christoffel c{bd} \right),
    \label{eq:defScalarCurvature}
\end{align}
where 
\begin{align}
    \christoffel c{ab} = \frac 12 g^{cd}(g_{da,b}+g_{db,a}-g_{ab,d})
\end{align}
are the \textit{Christoffel symbols of the first kind}. In the equations above, note that we have used Einstein's summation convention and that the commas in the subscripts indicate a partial derivative: for example, $\christoffel c{ab,d} = \partial_d \christoffel c{ab}$.  For more details on the scalar curvature, we refer the reader to \cite{Carroll:2004st}.

Apart from its importance in areas in differential geometry and in cosmology, the Ricci scalar is slowly finding applications in hitherto unexpected places for example the study of phase-transitions in quantum many-body systems  \cite{Dey2012, rizaerdem2020, michaelkolodrubetzvladimirgritsevanatolipolkovnikov2013} and in classical machine learning, a formulation of neural nets in the context of Riemannian Geometry has been explored \cite{hauser2017principles} while specific use of Riemannian curvature has been explored in \cite{8546273, 8746812}. 

 Using the isomorphism $HP^1 \simeq S^4$ where $HP^1$ is the quaternionic projective space we can incorporate the concurrence as part of the calculation of the \textit{Mannoury-Fubini-Study} metric \cite{levay2004geometry} as follows:

\begin{equation}
    \label{mannoury-fubini-study}
    g_{\mu \nu} = \frac{1}{1- C^2}dC^2 + C^2 d\chi^2 + (1-C^2)(d \Phi^2 + \sin^2 \Theta d \Theta^2) 
\end{equation}
where $w = |C| e^{i \chi}$ (defined in Sec. $2.3$), and $0 < \Phi \leq 2 \pi,  0 <  \Theta \leq \pi $ are the usual polar co-ordinates for $B^3$ (three dimensional ball). We may make an amusing observation that this is the same metric as  the Euclidean Schwarzshild metric for fixed time.

The metrics for the four circuits are therefore

\begin{align}
   \label{eq:pqc_metric1}
    g_{\text{HEA}} &= \begin{pmatrix} 
       \frac{1}{1 - C^2_{\text{HEA}}} & 0 & 0 & 0 \\
       0 & C^2_{\text{HEA}}& 0 & 0 \\
       0 & 0 & (1- C^2_{\text{HEA}}) & 0 \\
       0 & 0 & 0 & (1- C^2_{\text{HEA}})
    \end{pmatrix}  \\
    \label{eq:pqc_metric2}
    g_{\text{LDCA}} &=  \begin{pmatrix} 
       \frac{1}{1 - C^2_{\text{LDCA}}} & 0 & 0 & 0 \\
       0 & C^2_{\text{LDCA}}& 0 & 0 \\
       0 & 0 & (1- C^2_{\text{LDCA}}) & 0 \\
       0 & 0 & 0 & (1- C^2_{\text{LDCA}})
    \end{pmatrix}  \\
    \label{eq:pqc_metric3}
      g_{\text{QGAN}} &=  \begin{pmatrix} 
       \frac{1}{1 - C^2_{\text{QGAN}}} & 0 & 0 & 0 \\
       0 & C^2_{\text{QGAN}}& 0 & 0 \\
       0 & 0 & (1- C^2_{\text{QGAN}}) & 0 \\
       0 & 0 & 0 & (1- C^2_{\text{QGAN}}) \sin \Theta_{\text{QGAN}}
    \end{pmatrix} \\
    \label{eq:pqc_metric4}
    g_{\text{sHEA}} &=  \begin{pmatrix} 
       \frac{1}{1 - C^2_{\text{sHEA}}} & 0 & 0 & 0 \\
       0 & C^2_{\text{sHEA}}& 0 & 0 \\
       0 & 0 & (1- C^2_{\text{sHEA}}) & 0 \\
       0 & 0 & 0 & (1- C^2_{\text{sHEA}}) \sin \Theta_{\text{sHEA}}
    \end{pmatrix}
\end{align}
where:
\begin{equation}
\sin \Theta_{\text{QGAN}} =  \sqrt{1 - (\sin \theta_1 (\sin \theta_3 \cos \theta_5) + \sin \theta_5 \cos \theta_2 \cos \theta_3))^2}
\end{equation}
for the QGAN ansatz and
\begin{equation}
    \sin \Theta_{\text{sHEA}} =   \sqrt{1 - y^2},
\end{equation}
where 
\begin{align}
    y &= 
    \frac{1}{2} \bigg(\sin (\theta _1) \cos \bigg(\frac{\theta _3}{2}\bigg) \bigg(-\bigg(-\cos (\theta _2)+\bigg(\cos (\theta _2)+1\bigg) \cos \bigg(\frac{\theta _4}{4}\bigg)+1\bigg) \cos (\theta _5) \nonumber\\ 
    &\quad -\sin \bigg(\frac{\theta _4}{4}\bigg) \sin (\theta _5) \bigg(\cos (\theta _2)+1\bigg)\bigg)-\sin(\theta _2) \sin \bigg(\frac{\theta _3}{2}\bigg) \bigg(2 \sin \bigg(\frac{\theta _4}{4}\bigg) \sin ^2\bigg(\frac{\theta _1}{2}\bigg) \cos (\theta _5)\nonumber\\ &\quad +\sin (\theta _5) \bigg(\cos (\theta _1) +\big(\cos (\theta _1)-1\big) \cos \bigg(\frac{\theta _4}{4}\bigg)+1\bigg)\bigg)\bigg)
\end{align}
for the sHEA ansatz.

The scalar curvatures $\mathcal R_{\mathcal A}$ (see Eq.~\eqref{eq:defScalarCurvature}) of all the PQC metrics (Eqs.~\eqref{eq:pqc_metric1}--\eqref{eq:pqc_metric4}) $g_{\mathcal A}$ (for $\mathcal A = $ HEA, LDCA, QGAN, sHEA) are calculated to be:
\begin{align}
    \label{eq:ricci_scalar}
    \mathcal R_{\mathcal A} = \frac{2(6C_{\mathcal A}^2-5)}{C_{\mathcal A}^2-1}.
\end{align}
Note that the Ricci scalar can be calculated by just knowing the metric. Since the metric in our case is just a function of the concurrence of a general 2-qubit circuit, Eq.~\eqref{eq:ricci_scalar} is completely general for the two-qubit case.

We note that the formula is a simple rational function of the concurrence, and from the formula, we can see that singularities on the base manifold correspond to maximally entangled states with concurrence value of one. 

\begin{figure}[ht]
\centering
\includegraphics[scale=0.22]{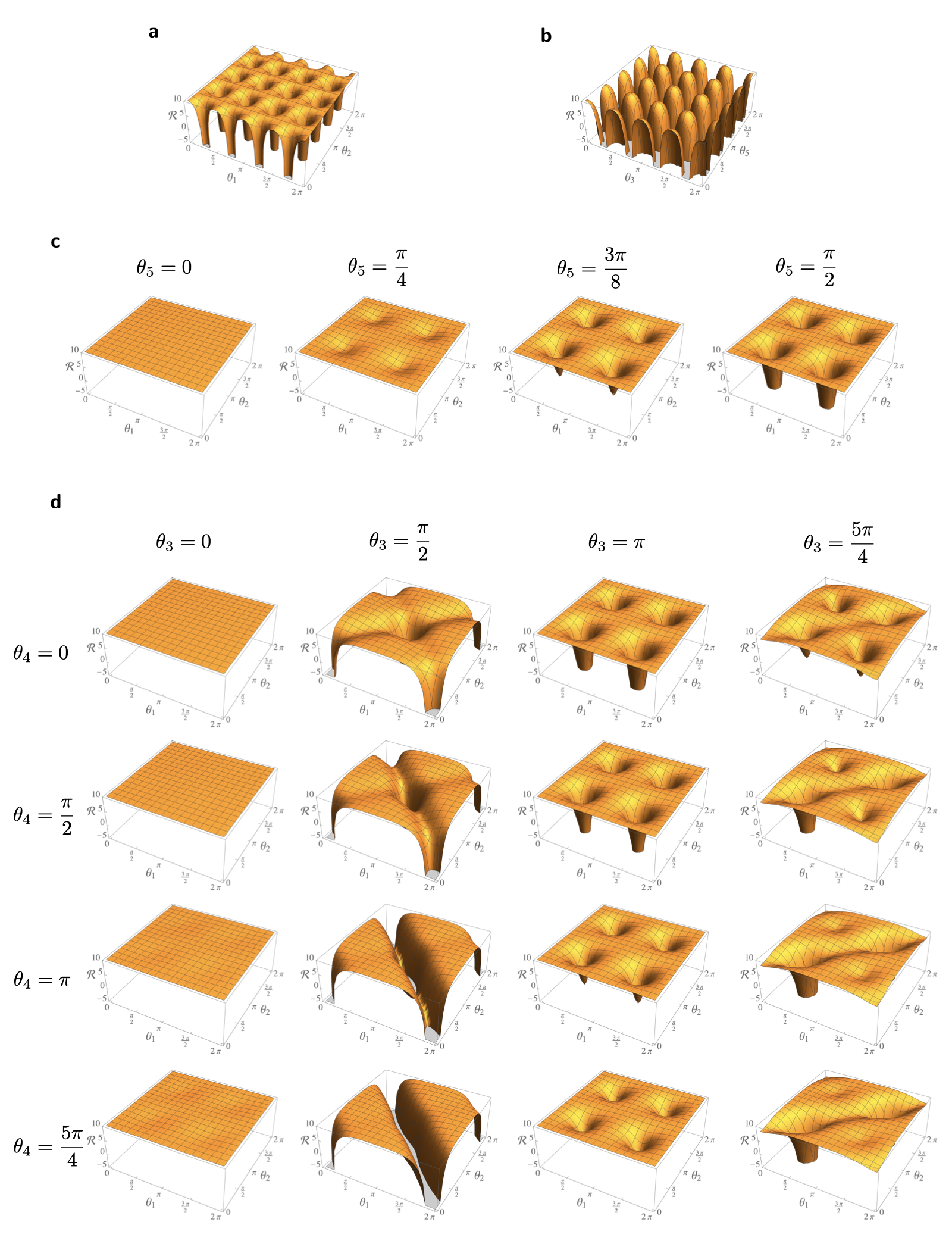}
\caption{
Scalar curvature ``landscapes'' of the four circuit blocks: (a) HEA, (b) LDCA, (c) QGAN, and (d) sHEA.
In the case of (a) and (b), the scalar curvature (denoted using $\mathcal{R}$) is a function of only two parameters: $\theta_1$ and $\theta_2$ for HEA and $\theta_3$ and $\theta_5$ for LDCA.
For the other two circuit blocks, in which $\mathcal{R}$ is a function of more than two circuit parameters, $\mathcal{R}$ is plotted as a function of $\theta_1$ and $\theta_2$, while scanning specific values of the remaining parameters to produce snapshots of the landscapes.
We limit the range of $\mathcal{R}$ to [-5, 10] for ease of visualization.
}
\label{fig:curvature_landscape}
\end{figure}

The scalar curvature of each of the circuit written in terms of their circuit parameters are:
\begin{align}
    \mathcal R_{\text{HEA}}  &= 12 -\frac{1}{\sin (2 \text{$\theta_1$}) \cos (2 \text{$\theta_2$})+1}+\frac{1}{\sin (2 \text{$\theta_1$}) \cos (2 \text{$\theta_2$})-1}, \label{eq:hea_ricci_scalar}\\
    \mathcal R_{\text{LDCA}} &= 12 - 2 \sec ^2(2 \text{$\theta_3$}) \sec ^2(2 \text{$\theta_5$}), \label{eq:ldca_ricci_scalar}\\
    \mathcal R_{\text{QGAN}} &= 12 - \frac{1}{\sin (\text{$\theta_1$}) \sin (\text{$\theta_2$}) \sin (\text{$\theta_5$})+1}+\frac{1}{\sin (\text{$\theta_1$}) \sin (\text{$\theta_2$}) \sin (\text{$\theta_5$})-1}, \label{eq:qgan_ricci_scalar}\\
    \mathcal R_{\text{sHEA}} &= 
    \frac{
    \Bigg( \splitfrac{
    12 \sin ^2\left(\theta _1\right) \sin ^2\left(\theta _2\right) \left(\cos \left(\theta _3\right)-\cos \left(\frac{\theta _4}{4}\right)\right){}^2
    }{
    + 12\left(\sin \left(\theta _3\right) \left(\cos \left(\theta _1\right) \cos \left(\theta _2\right)+1\right)-\sin \left(\theta _1\right) \sin \left(\theta _2\right) \sin \left(\frac{\theta _4}{4}\right)\right){}^2 - 40
    }
    \Bigg)
}{
    \splitfrac{
    \sin ^2\left(\theta _1\right) \sin ^2\left(\theta _2\right) \left(\cos \left(\theta _3\right)-\cos \left(\frac{\theta _4}{4}\right)\right){}^2
    }{
    +\left(\sin \left(\theta _3\right) \left(\cos \left(\theta _1\right) \cos \left(\theta _2\right)+1\right)-\sin \left(\theta _1\right) \sin \left(\theta _2\right) \sin \left(\frac{\theta _4}{4}\right)\right){}^2-4.
    }
}.
\label{eq:shea_ricci_scalar}
\end{align}
These scalar curvatures are additionally plotted in Fig. \ref{fig:curvature_landscape}. 
For HEA and LDCA, their Ricci scalars are each a function of two circuit parameters, i.e. only two parameters influence the entanglement in the system. 
On the other hand, the Ricci scalars of QGAN and sHEA are functions of three and four parameters, respectively. 
To visualize how parameter values impact the curvature, values of $\theta_1$ and $\theta_2$ are scanned over range $[0, 2\pi]$ while values of other parameters, if they appear in the expression of the Ricci scalar, are fixed at particular values.
With QGAN, by tuning the value of $\theta_5$ from $0$ to $\frac{\pi}{2}$, we observe a gradual emergence of ``wells'' of negative curvature.
With sHEA, the landscape depends significantly on values of $\theta_3$ and $\theta_4$.
For instance, when $\theta_3 = \pi$, there are wells of negative curvature similar to those from QGAN's landscape.
However, when $\theta_3$ is decreased to approximately $\frac{\pi}{2}$, the wells are replaced by ``valleys'' of negative curvature.
Comparing these landscapes provides some insight into the (relative) entangling capabilities of the circuit blocks and the ease of generating high entanglement states by tuning the circuit parameters.
Circuits corresponding to curvature landscapes with extensive or large regions of negative curvatures (i.e. high concurrences) such as LDCA, are able to more readily generate states with high entanglement compared to circuits with landscapes with limited regions of negative curvature similar to that of QGAN.
Additionally, it appears easier to generate high entanglement states with LDCA than to do so using circuits such as QGAN and sHEA, in which one must tune multiple circuit parameters, several of which need to be set near specific values to generate states with high entanglement.

Why might one be interested in calculating the Ricci scalar? After all, from the face of it, it seems to contain the same amount of information as the concurrence. 
The answer lies in the goal of this study, namely understanding PQCs from a geometric point of view and their optimization performance with regards to the Quantum Natural Gradient. 
We have the intuition that the Quantum Natural Gradient somehow incorporates the geometric information of the projective Hilbert space in the optimization procedure in order improve the speed of convergence. The question is that can we map out the geometry of the quantum circuits and see \textit{geometrically} what the quantum natural gradient is doing differently? This is the question we take up in next section and explore.

\section{Insights into VQE performance through scalar curvature: a two-qubit study}\label{sec:numerical_experiments}

In this section, we numerically demonstrate the connection between the geometry and performance of two-qubit parameterized quantum circuits.
To quantify the performance, we consider a toy problem instance of the Variational Quantum Eigensolver (VQE) algorithm \cite{Peruzzo2014} which, despite its simplicity, shows how the scalar curvature of a parameterized quantum circuit for two qubits may help inform the effectiveness of an ansatz prior to execution of the algorithm.
Namely, the task-at-hand is to estimate the ground state energy of molecular hydrogen at a particular bond length, employing each of the two-qubit circuits.
We consider the two-qubit Hamiltonian for molecular hydrogen from an early VQE experimental study \cite{O_Malley2016}:
\begin{equation}
    H = \nu_1 \mathbb{I} + \nu_2 Z_1 + \nu_3 Z_2 + \nu_4 Z_1 Z_2 + \nu_5 X_1 X_2 + \nu_6 Y_1 Y_2,
\end{equation}
where $\boldsymbol \nu = \{ \nu_1, \nu_2, ..., \nu_6 \}$ corresponds to Hamiltonian coefficients.
For this problem, the ground state wave function is of the form: $\ket{\psi_g} = \alpha(r_{H-H}) \ket{01} + \beta(r_{H-H}) \ket{10}$, where the wave function coefficients $\alpha$ and $\beta$ are functions of $r_{H-H}$, the inter-atomic distance between the two Hydrogen atoms.
At $r_{H-H} = 3.19 \ \angstrom$, the ground state wave function is a highly entangled state of the form:
\begin{equation}
    \ket{\psi_g} = \alpha \ket{01} + \beta \ket{10},
\end{equation}
for $\alpha, \beta \in \mathbb{C}$ where $|\alpha|^2 \approx 0.47$ and $|\beta|^2 \approx 0.53$. 
In this case, the ability to generate (highly) entangled states using a parameterized quantum circuit is  important for representing the solution state.
We executed wave function simulations for the VQE calculations. For the optimization in VQE, we consider the standard gradient descent optimizer as well as the Quantum Natural Gradient optimizer using the block-diagonal and diagonal matrix approximations for the Fubini-Study metric tensor \cite{Stokes2019}.
We fix the step size of each type of gradient descent to $0.05$ and terminate each optimization run based on a convergence threshold of $10^{-6}$.

Using this toy problem, we discuss two main observations:\footnote{We repeat the VQE simulations and analyses for a bond length corresponding to an unentangled ground state and obtain similar results. These results are shown in Appendix \ref{app:50_runs_product}.}
\begin{enumerate}
    \item Particular circuit structures lead to precise and/or accurate solutions. Using 50 independent optimizations, we show that circuits like LDCA initialize states at high concurrences and consistently and rapidly converge to the solution. Others have wider spread in the final energies and their accuracies. We argue that further insight about these results can be understood from the scalar curvature.
    \item We investigate the role of the fiber space by considering the QGAN circuit as a case study. This circuit, as constructed, is unable to reach the ground state but is able to do so by appending local rotation gates. By calculating the curvature, we are also able to see how the Quantum Natural Gradient is able to take advantage of the geometry while the standard gradient descent does not.
\end{enumerate}

\subsection{Curvature landscape and optimization}
\begin{figure*}[ht]
\centering
\includegraphics[scale=0.34]{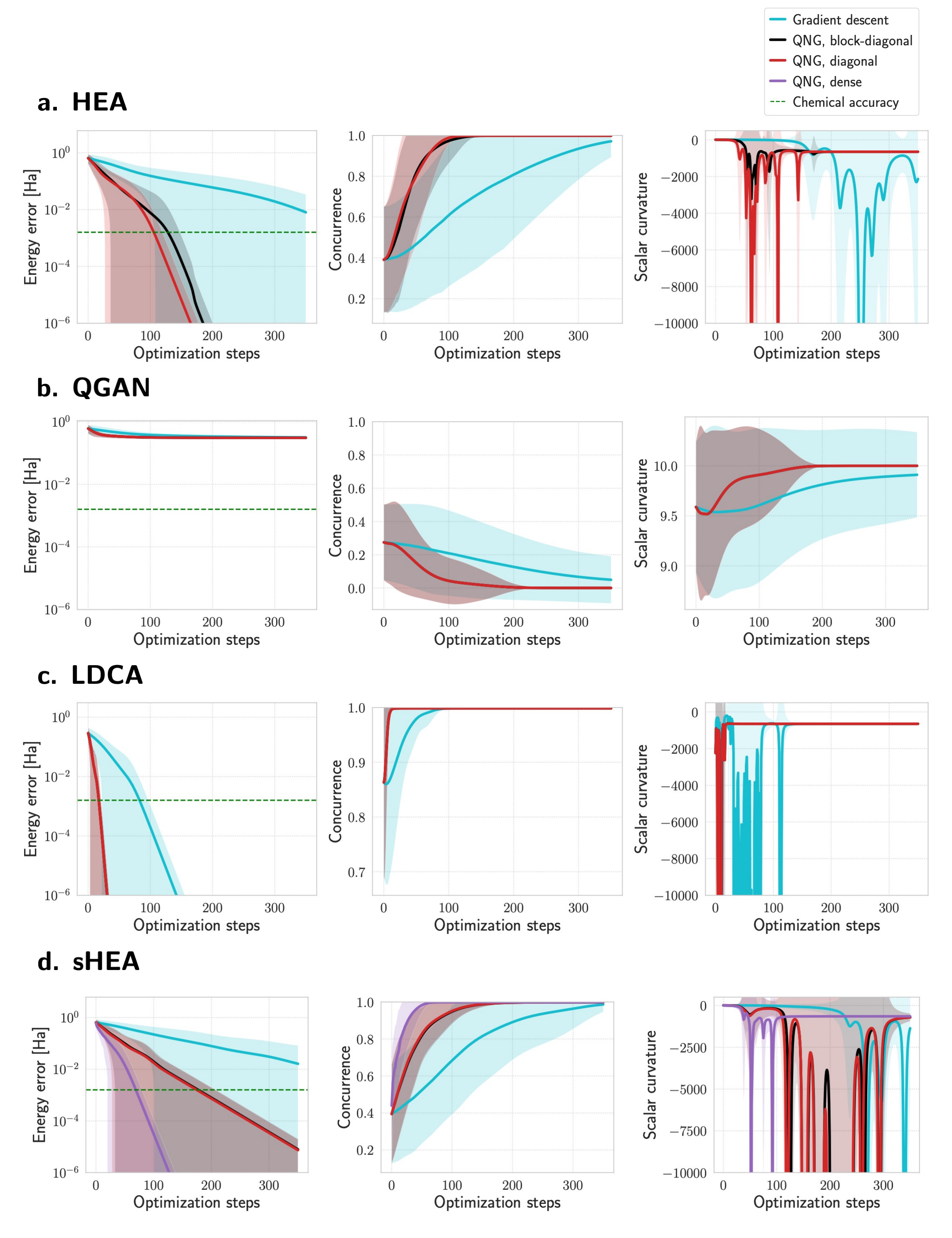}
\caption{
VQE results over 50 random initial points using the four two-qubit ansatze for estimating the ground state energy of molecular hydrogen at $R_{\text{H-H}} = 3.19 \angstrom$.
Ground state at this bond length is highly entangled.
Solid lines show the average values of energy error, concurrence, and scalar curvature.}
\label{fig:h2_entangled_ground_state_data}
\end{figure*}

We first investigate the connection between the curvature landscape and VQE optimization. 
Fig.~\ref{fig:h2_entangled_ground_state_data} tracks the energy error, concurrence, and scalar curvature of optimization paths using the four circuit blocks.
For each circuit, 50 independent optimization trials were performed using random parameter initialization. 
For three of the four circuits, namely HEA, LDCA, and sHEA, using Quantum Natural Gradients enabled the optimizations to converge to errors below the chemical accuracy threshold ($\approx 10^{-3}$ Ha) within 200 descent steps.
Referring back to Fig.~\ref{fig:curvature_landscape}, these three circuits correspond to curvature landscapes with extensive regions of negative curvature or high concurrence.
For sHEA, $\theta_3$ and $\theta_4$ were tuned over the course of each optimization such that highly entangled states became accessible (e.g. $\theta_3 \approx \pi/2$).

In particular, optimization runs for LDCA started at higher values of concurrence on average and rapidly converged to accurate ground state energies within around 20 descent steps for QNG methods. In addition, the standard deviation over 50 runs is smaller than those corresponding to other circuits.

Additional insights into the optimization procedure can be gathered by looking at the curvature landscapes and also the scalar curvatures reached during the optimization process.
\begin{enumerate}[label=(\roman*)]
    \item For the circuits that reach the ground state, we observe more regions of negative curvature; these regions in the Hilbert space are useful in accelerating the optimizations. Having more of these regions seems to correlate with performance of the optimization process. Consider the curvature landscape of LDCA, in which we observe that there are small, repeating hills of positive curvature that are surrounded by regions of negative curvature. This implies that LDCA may have a greater access to highly entangled states that are additionally constrained to a subspace spanned by $\{\ket{01}, \ket{10}\}$, both of which led to superior performance in the VQE algorithm instance. On the other hand, if one looks at the QGAN curvature landscapes, we see that for majority of parameter settings the curvature is positive.
    \item Because the Quantum Natural Gradient is tailored for the wave function geometry, we see that it is able to reach these areas of negative curvature faster than standard gradient descent. Once it reaches these regions, the optimization process is sped up earlier on in the process allowing for a lower number of iterations. In all the cases looked at, the standard gradient descent takes longer to find these regions.
    \item Lastly, we briefly comment on the relative costs of Quantum Natural Gradient, which depend on the number of non-zero elements in the metric tensor or its approximations.
    From Eqs.~\ref{eq:fs_metric_tensor_hea}- \ref{eq:fs_metric_tensor_shea}, we see that the diagonal or block-diagonal approximation of the metric tensor for LDCA captures all of the non-zero elements. 
    This is in contrast to the other circuits which have at least two unique elements that are not captured by the approximations to the metric tensor.
    To make better use of QNG, these other circuits may require computations of elements that are not captured by the block-diagonal or diagonal approximations.
    For example, in Fig. \ref{fig:h2_entangled_ground_state_data}d, we ran the VQE optimizations using the dense or full metric tensor for sHEA, a case in which the block-diagonal approximation does not capture many of the non-zero elements.
    Shown using purple lines, we observe that the optimization significantly improves in efficiency though at the cost of more function calls needed at each QNG descent step to compute the full metric tensor.
    On the other hand, LDCA only requires one non-zero element of the metric tensor to be computed at each QNG descent step.
    This appears to imply that certain circuits, by the way that they are parameterized, are better suited for QNG methods than others.
\end{enumerate}

\subsection{Non-trivial role of adding single-qubit gates}
\begin{figure}[ht]
\centering
\includegraphics[scale=0.33]{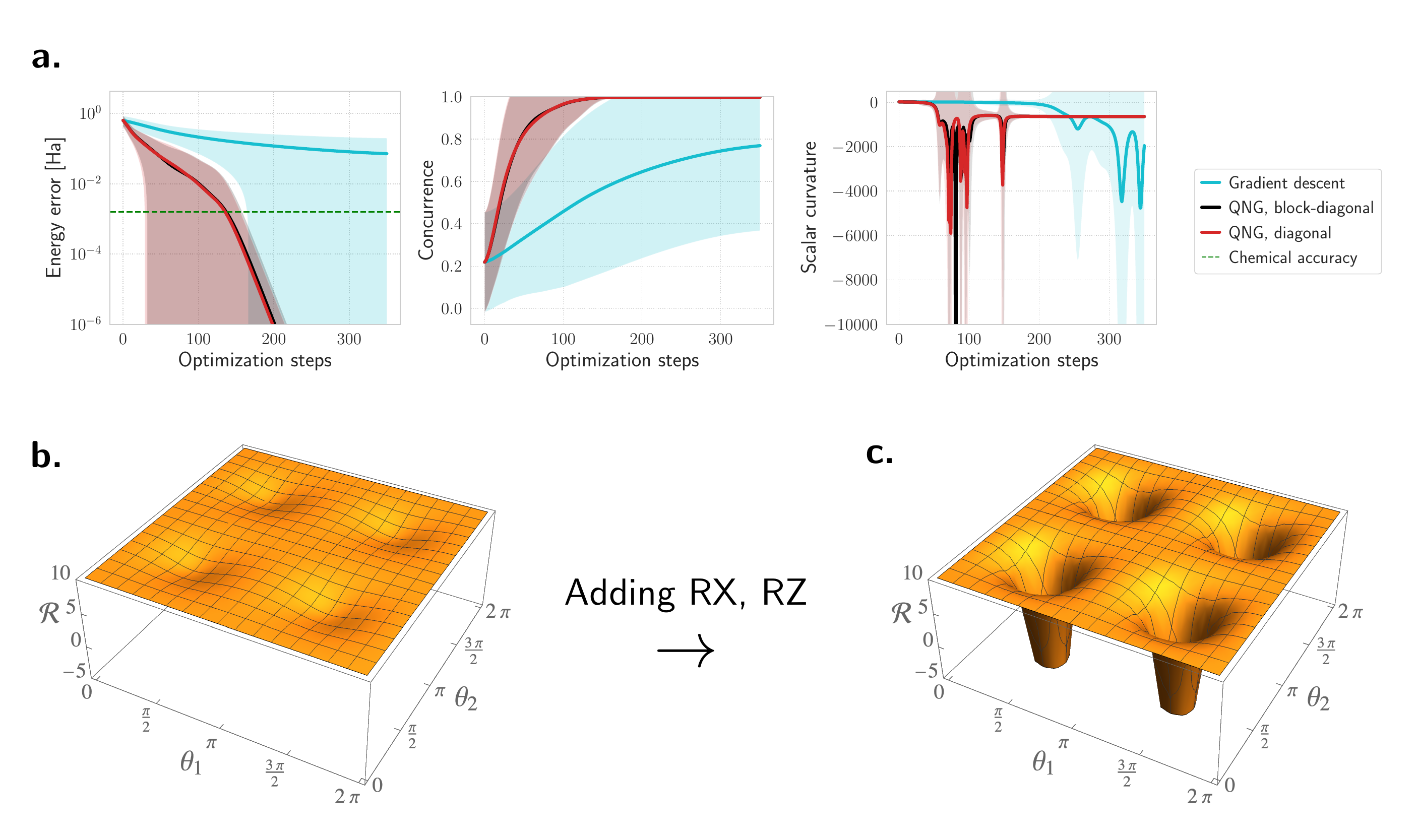}
\caption{
(a) Energy error, concurrence, and curvature tracked over 50 independent optimization trials of QGAN augmented with local rotations (RX and RZ applied to each qubit). (b) Curvature landscape corresponding to final/optimized parameter value of $\theta_5$ of a particular optimization run using the original QGAN circuit. (c) Curvature landscape corresponding to final parameter value of $\theta_5$ of a particular optimization run using \emph{augmented} QGAN circuit.}
\label{fig:qgan_aug}
\end{figure}

As observed in Fig. \ref{fig:h2_entangled_ground_state_data}b, QGAN consistently fails to reach the ground state. 
QGAN, as constructed, is unable to reach states corresponding to high entanglement, or equivalently negative curvature.
Often, in similar situations in which the circuit seems insufficient, one adds a set of gates or a circuit layer in the hopes of providing greater flexibility to reach the solution state.
Thus, we tried augmenting the QGAN circuit with local rotations, adding $RX$ then $RZ$ single-qubit gates to each of the two qubits. This adds four new parameters to the circuit.
Appending local rotations should not impact the concurrence and thus the curvature. 
It was surprising to observe, however, that adding local rotations greatly improved the optimization as shown in Fig. \ref{fig:qgan_aug}a. 
While the average value of concurrence at the start of the optimization using the augmented QGAN was lower than that corresponding to the original QGAN circuit, the updated QGAN circuit was able to estimate the ground state energy with high accuracy using quantum natural gradient methods.
The puzzle is then the following: while the concurrence expression in Eq.~\eqref{eq:qgan_concurrence} should not change for the augmented QGAN (since single qubit gates should not increase entanglement), the additional parameterized single-qubit gates somehow appear to have aided in guiding the other circuit parameters to values such that highly entangled states are accessible (shown in Fig. \ref{fig:qgan_aug}c). In other words, without increasing the entanglement in the circuit, we are able to turn an optimization procedure that does not get to an entangled state to one that does. We suspect that this is where another part of the geometry plays a part, namely the fibers. Locally, we have the geometry being $S^4 \times S^3$ but this is not true globally, which means that how we move through the fibers has a non-trivial impact on the optimization process. 
This case also illustrates the role of geometry plays in the optimization process because the standard gradient descent still fails to find the ground state, at least within a reasonable number of optimization steps. 
From the point of view of just the optimization process, we simply see the concurrence fails to reach the target value and the energy fails to reach low enough values, but as we look at what happens to the scalar curvature, we see the Quantum Natural Gradient descent finds pockets of high negative curvature allowing for the optimization path to more easily move. In other words, as we change parts of the available geometry, the Quantum Natural Gradient is able to better leverage that.

\section{Concluding remarks}\label{sec:conclusion}

Often, investigations or uses of PQCs are limited to considering their inputs (parameters) and outputs (resulting quantum states or observables).
Our work provided an investigation into the inner workings of two-qubit PQCs, which we argue are the simplest instances of PQCs and yet can still provide valuable insights for better understanding these circuits.
We first defined PQCs from a geometric perspective. With the help of specifically two-qubit geometry, we explicitly worked out the intrinsic co-ordinates for four examples of two-qubit PQC blocks, at least for the base manifold, which is important for characterizing entanglement in the circuit. 
As a consequence of our ability to parameterize our circuits in terms of two-qubit geometry, we introduced a notion of local geometric expressibility, which describes how much of the two-qubit geometry can be explored by a two-qubit PQC. 

In trying to understand the connection between the geometry of the projective Hilbert space and how it is used by the Quantum Natural Gradient in the optimization process, we used the Mannoury-Fubini-Study metric as a way to calculate the curvature of the base manifold, $S^4$. 
This provided a simple and remarkable connection between the curvature of the base manifold and the amount of entanglement in the ansatz.

With this connection we were able to establish a bridge between the amount of entanglement and quality of the optimization procedure to the geometry of the projective Hilbert Space. This allowed us to notice a correlation between the ability of the ansatz to find earlier on in the optimization process regions of high negative curvature and acceleration of the optimization process. We strongly suspect that this connection is not merely a correlation but represents a chain of causation, for this we give two pieces of evidence:
\begin{enumerate}[label=(\roman*)]
    \item The performance of the QGAN circuit which could not find either the entangled ground state or the product ground state was enhanced by creating the augmented QGAN ansatz i.e. by adding single qubit rotations. The optimization process was able to reach regions of high negative curvature that were not accessible before and after which the circuit was able to reach the entangled ground state. 
    Furthermore, in the numerical simulations for which the ground state is a product state and hence lies in a region of positive curvature (Appendix \ref{app:50_runs_product}), the augmented QGAN performed better by initially finding regions that are less positive than the original QGAN.
    
    \item Secondly, by inspecting the performance of standard gradient descent we see that for the circuits that did find the ground state, the regions of high negative curvature were found later in the optimization process than in the case for QNG. Indeed the behavior of the standard gradient descent for the augmented QGAN we believe is highly suggestive. The standard gradient descent in the augmented QGAN ansatz is not able to find these regions of high negative curvature and is not able to reach the ground state for a circuit we know can find the ground state. On the other extreme end, we can look at the performance of sHEA and consider the performance of the dense QNG which finds region of negative curvature the earliest, one sees that this correlates with finding the ground state much more efficiently than the other approximations of QNG and the standard gradient descent.
\end{enumerate}

In summary, the geometric perspective through the use of the scalar curvature has provided us with a tool to explain why, for example, why LDCA outperforms other circuits other than just explaining via tracking energies or entanglement.
Overall, we show that the number of parameters of a given PQC does not necessarily correspond to high circuit flexibility or capability; it matters \emph{how} the circuit is parameterized. In connection with this point, we observed how single qubit gates can significantly impact the circuit performance,
i.e. do more than just provide extra parameters. 
The single qubit gates allow us to explore the fibers of the geometry in such ways as to find regions of negative or less positive curvature. This effect is ultimately possible because the geometry is not just a cartesian product of the base manifold and the fibers but some complicated twist.
The geometric perspective also provides insights into how and why the QNG is better than the standard gradient descent; the QNG can find the regions of negative curvature, and if there are none, it opts for regions with less positive curvature early in the optimization process. We stress that the geometric perspective is necessary in order to understand the QNG since the QNG does not depend on the cost function.

While we provided a extensive study on two-qubit PQCs, there remain several puzzles or open questions, which we outline in the following subsection for future work.

\subsection{Future directions and open questions}
\label{sec:future_directions}

\subsubsection*{Decomposition Problem}
Refs.~\cite{Stokes2019, Yamamoto2019} numerically showed the potential for Quantum Natural Gradients in optimizing parameterized quantum circuits in the context of VQE. 
A main challenge will be to scale up this method for larger and deeper quantum circuits. In an effort to formulate a hopefully simpler scenario that may be able to be scaled, we describe the following ``decomposition problem'':

Suppose we have a four-qubit circuit, the first and second qubits are entangled using one of the PQC building blocks, e.g. LDCA block while the third and fourth qubits are also entangled using a two-qubit block.
Overall, we have 2 two-qubit subsystems, each of which we know the metric tensor, call them $g^{(12)}$ and $g^{(34)}$.
However, suppose we add a static/non-parametric entangler (e.g. CNOT) to the second and third qubits. 
How is the metric tensor of the four-qubit system $g^{(1234)}$ constructed, compared to structures of $g^{(12)}$ and $g^{(34)}$? The idea being, rather than anaylzing the $n$ qubit geometry from the ground up in order to understand PQCs for $n$ qubits, can we use the simpler geometry of 2 qubits to bootstrap our way to understanding higher qubit number PQCs?

\subsubsection*{Three-qubit case}
For this work, we crucially relied on the fact that the geometry of 2 qubits could easily be understood since it was simply the Hopf fibration $S^7$. As a consequence of Hopf Invariant One theorem \cite{johnfrankadams1958}, there is one more Hopf fibration that can be studied in great detail namely the Hopf fibration $S^7 \rightarrow S^{15} \xrightarrow{\pi} S^8 $. In this case, the use of octonions should be helpful in parameterizing the geometry.

\section*{Acknowledgements}
All the authors would like to thank their colleagues at Zapata Computing for their feedback. 
The metric tensors and scalar curvatures for each PQC were computed using EinsteinPy \cite{einsteinpy} and Mathematica \cite{Mathematica}.
The VQE simulations were implemented using OpenFermion \cite{openfermion} and PennyLane \cite{Bergholm2018}. DEK acknowledges funding support from the National Research Foundation, Singapore, through Grant NRF2021-QEP2-02-P03.

\appendix
\section{Gate definitions}\label{app:gate_definitions}
We define the rotation operations in the following way:
\begin{align}
    R_P(\theta) = \exp ( -i \theta P /2 ),
\end{align}

where $P \in \{ X, Y, Z\}$ for single-qubit gates and $P \in \{ XX, YY, ZZ, XY, YX\}$ for two-qubit gates.

We define the iSWAP$^\dagger$ gate as:
\begin{equation}
    \text{iSWAP}(\theta)^\dagger = \text{e}^{-i \theta (X \otimes X + Y \otimes Y) /2},
\end{equation}

and the CPHASE gate as:
\begin{equation}
    \text{CPHASE}(\phi) = \text{e}^{-i \phi (I-Z)\otimes (I-Z) /4}.
\end{equation}

{
\renewcommand{\arraystretch}{1.5}
\begin{table}[ht]
  \centering
  \caption{Abbreviations and symbols}
  \begin{tabular}{ll}
    \toprule
    \textbf{Symbol/Abbreviation} & \textbf{Description}\\
    \midrule
    $G$, $G(\boldsymbol\theta)$ & Fubini-Study metric tensor\\
    $\mathcal{R}$ & Scalar curvature\\
    $C$, $C(\psi)$ & Concurrence\\
    $\boldsymbol\theta$ & Parameter vector\\
    $\theta_j$ & $j$-th parameter\\
    PQC & Parameterized quantum circuit\\
    HEA & Hardware-efficient ansatz\\
    LDCA & Low-depth circuit ansatz \cite{Dallaire-Demers2018}\\
    QGAN & Quantum generative adversarial network\\
    \bottomrule
  \end{tabular}
  \label{tab:table}
\end{table}
}

\section{Hopf base and fiber parameters for various circuit blocks}
\label{app:hopfBaseFiber}
\subsection{Hardware-efficient ansatz (HEA)}

The output state of the quantum hardware-efficient ansatz (HEA) depicted by Fig. \ref{fig:circuit_diagrams_v2}(a) is
\begin{align}
 \ket{\psi}_{HEA} &=    \cos \left(\theta _1\right) \cos \left(\theta _3\right) \cos \left(\theta _2+\theta _4\right)-\sin \left(\theta _1\right) \sin \left(\theta _3\right) \sin \left(\theta _2-\theta _4\right) 
\ket{00}\nonumber\\
&\quad
+\sin \left(\theta _2+\theta _4\right) \cos \left(\theta _1\right) \cos \left(\theta _3\right)-\sin \left(\theta _1\right) \sin \left(\theta _3\right) \cos \left(\theta _2-\theta _4\right)
\ket{01} 
\nonumber\\
&\quad
+\sin \left(\theta _3\right) \cos \left(\theta _1\right) \cos \left(\theta _2+\theta _4\right)+\sin \left(\theta _1\right) \sin \left(\theta _2-\theta _4\right) \cos \left(\theta _3\right)
\ket{10} 
\nonumber\\
&\quad
+\sin \left(\theta _1\right) \cos \left(\theta _3\right) \cos \left(\theta _2-\theta _4\right)+\sin \left(\theta _3\right) \sin \left(\theta _2+\theta _4\right) \cos \left(\theta _1\right)
   \ket{11}.
\end{align}

The angle parameters of the $S^4$ Hopf base $(\theta_A,\phi_A)$ and $(\chi,\xi)$ and the equivalent Cartesian coordinates $(x_0,x_1,x_2,x_3,x_4)$ are calculated to be
\begin{align}
    x_0 &= \cos \left(2 \theta _1\right) \cos \left(2 \theta _3\right)-8 \sin \left(\theta _1\right) \sin \left(\theta _2\right) \sin \left(\theta _3\right) \cos \left(\theta _1\right) \cos \left(\theta _2\right) \cos \left(\theta _3\right)\\
    x_1 &= \sin \left(2 \theta _1\right) \sin \left(2 \theta _2\right) \cos \left(2 \theta _3\right)+\sin \left(2 \theta _3\right) \cos \left(2 \theta _1\right) \\
x_2 &= 0 \\
x_3 &= \sin \left(2 \theta _1\right) \cos \left(2 \theta _2\right) \\
x_4 &= 0
\end{align}
and
\begin{align}
    \theta_A &= \arccos\left(\cos \left(2 \theta _1\right) \cos \left(2 \theta _3\right)-8 \sin \left(\theta _1\right) \sin \left(\theta _2\right) \sin \left(\theta _3\right) \cos \left(\theta _1\right) \cos \left(\theta _2\right) \cos \left(\theta _3\right)\right) \\
    \phi_A &= \arccos\left(\frac{\sin \left(2 \theta _1\right) \sin \left(2 \theta _2\right) \cos \left(2 \theta _3\right)+\sin \left(2 \theta _3\right) \cos \left(2 \theta _1\right)}{\sqrt{1-\left(\cos \left(2 \theta _1\right) \cos \left(2 \theta _3\right)-8 \sin \left(\theta _1\right) \sin \left(\theta _2\right) \sin \left(\theta _3\right) \cos \left(\theta _1\right) \cos \left(\theta _2\right) \cos \left(\theta _3\right)\right){}^2}}\right) \\
    \chi &= \frac{\pi }{2}
    \\
    \xi &= \frac{\pi }{2}.
\end{align}

The quaternion $q_\pm=\langle c_\pm|\psi_H\rangle$ is calculated to be
\begin{align}
    \label{hea_fiber}
    q_\pm = \frac 1{2\lambda_2} (e_\pm + jf_\pm)
\end{align}
where
\begin{align}
  e_\pm &= \left(\cos \left(\theta _1\right) \cos \left(\theta _3\right) \cos \left(\theta _2+\theta _4\right)-\sin \left(\theta _1\right) \sin \left(\theta _3\right) \sin \left(\theta _2-\theta _4\right)\right) \left(\lambda _3 \gamma_\mp+\lambda _1 \gamma_\pm\right) \\
  f_\pm &=
  \left(\sin \left(\theta _1\right) \sin \left(\theta _3\right) \cos \left(\theta _2-\theta _4\right)-\sin \left(\theta _2+\theta _4\right) \cos \left(\theta _1\right) \cos \left(\theta _3\right)\right) \left(-\lambda _3 \gamma_\mp-\lambda _1 \gamma_\pm \right)
\end{align}
where
\begin{align}
    \lambda_1 &=
    \sqrt{2 \sin \left(4 \theta _1\right) \sin \left(2 \theta _2\right) \sin \left(4 \theta _3\right)+4 \sin ^2\left(2 \theta _1\right) \left(\cos ^2\left(2 \theta _2\right)+\sin ^2\left(2 \theta _2\right) \cos ^2\left(2 \theta _3\right)\right)+4 \sin ^2\left(2 \theta _3\right) \cos ^2\left(2 \theta _1\right)}
    \nonumber\\
    \lambda_2
    &=
    \sqrt{\sin \left(4 \theta _1\right) \sin \left(2 \theta _2\right) \sin \left(4 \theta _3\right)+2 \sin ^2\left(2 \theta _1\right) \left(\cos ^2\left(2 \theta _2\right)+\sin ^2\left(2 \theta _2\right) \cos ^2\left(2 \theta _3\right)\right)+2 \sin ^2\left(2 \theta _3\right) \cos ^2\left(2 \theta _1\right)}
    \nonumber\\
    \lambda_3 &=
    2 \sin \left(2 \theta _1\right) \sin \left(2 \theta _2\right) \sin \left(2 \theta _3\right)-2 \cos \left(2 \theta _1\right) \cos \left(2 \theta _3\right)+2
    \nonumber\\
    \gamma_\pm &=
    \sqrt{1\pm \sqrt{1-\frac{1}{4} \sin ^2\left(2 \theta _1\right) \cos ^2\left(2 \theta
   _2\right)-\frac{1}{4} \left(\sin \left(2 \theta _1\right) \sin \left(2 \theta _2\right)
   \cos \left(2 \theta _3\right)+\sin \left(2 \theta _3\right) \cos \left(2 \theta
   _1\right)\right){}^2}}.
\end{align}

\subsection{Low-depth circuit ansatz (LDCA)}

The output state of the quantum hardware-efficient ansatz (HEA) depicted by Fig. \ref{fig:circuit_diagrams_v2}(a) is
\begin{align}
\ket{\Psi}_{LDCA} &=  e^{-\frac{1}{2} i \left(\theta _1-\theta _2-\theta _4\right)} \left(\cos \left(\theta _3\right) \cos \left(\theta _5\right)-i \sin \left(\theta _3\right) \sin \left(\theta _5\right)\right)
\ket{01} \nonumber\\
&\quad - e^{-\frac{1}{2} i \left(\theta _1-\theta _2-\theta _4\right)} \left(\sin \left(\theta _5\right) \cos \left(\theta _3\right)+i \sin \left(\theta _3\right) \cos \left(\theta _5\right)\right)
\ket{10} .
\end{align}

The angle parameters of the $S^4$ Hopf base $(\theta_A,\phi_A)$ and $(\chi,\xi)$ and the equivalent Cartesian coordinates $(x_0,x_1,x_2,x_3,x_4)$ are calculated to be
\begin{align}
    x_0 &= \cos \left(2 \theta _3\right) \cos \left(2 \theta _5\right)\\
    x_1 &= 0 \\
x_2 &= \sin \left(\theta _1-\theta _2-\theta _4\right) \sin \left(2 \theta _5\right)-\sin \left(2 \theta _3\right) \cos \left(\theta _1-\theta _2-\theta _4\right) \cos \left(2 \theta _5\right) \\
x_3 &= \sin \left(2 \theta _3\right) \sin \left(\theta _1-\theta _2-\theta _4\right) \cos \left(2 \theta _5\right)+\sin \left(2 \theta _5\right) \cos \left(\theta _1-\theta _2-\theta _4\right) \\
x_4 &= 0
\end{align}
and
\begin{align}
    \theta_A &= \arccos \left(\cos \left(2 \theta _3\right) \cos \left(2 \theta _5\right)\right) \\
    \phi_A &= \frac{\pi }{2} \\
    \chi &= \frac{\pi }{2}
    \\
    \xi &= -\arctan \left(\frac{\sin \left(2 \theta _3\right) \cos \left(2 \theta _5\right)+\sin \left(2 \theta _5\right) \cot \left(\theta _1-\theta _2-\theta _4\right)}{\sin \left(2 \theta _3\right) \cos \left(2 \theta _5\right) \cot \left(\theta _1-\theta _2-\theta _4\right)-\sin \left(2 \theta _5\right)}\right).
\end{align}

The quaternion $q_\pm=\langle c_\pm|\psi_H\rangle$ is calculated to be
\begin{align}
   \label{ldca_fiber}
    q_\pm = \frac 1{\sqrt {\mu_3}} (j g_\pm + k h_\pm),
\end{align}
where
\begin{align}
    g_\pm &= \left(\cos \left(\theta _3\right) \cos \left(\frac{1}{2} \left(\theta _1-\theta _2-\theta _4\right)\right) \cos \left(\theta _5\right)-\sin \left(\theta _3\right) \sin \left(\frac{1}{2} \left(\theta _1-\theta _2-\theta _4\right)\right) \sin \left(\theta _5\right)\right) (\mu_1 \gamma_\pm +\mu_2 \gamma_\mp ), \nonumber\\
    h_\pm &= \left(\sin \left(\frac{1}{2} \left(\theta _1-\theta _2-\theta _4\right)\right) \cos \left(\theta _3\right) \cos \left(\theta _5\right)+\sin \left(\theta _3\right) \sin \left(\theta _5\right) \cos \left(\frac{1}{2} \left(\theta _1-\theta _2-\theta _4\right)\right)\right) (-\mu_1 \gamma_\pm -\mu_2 \gamma_\mp ).
    \end{align}
    where
    \begin{align}
    \mu_1 &=
    \sqrt{-2 \cos \left(4 \theta _3\right) \cos ^2\left(2 \theta _5\right)-\cos \left(4 \theta _5\right)+3}, \nonumber\\
    \mu_2 &=
    2-2 \cos \left(2 \theta _3\right) \cos \left(2 \theta _5\right), \nonumber\\
\mu_3 &=    
    2 \left(-2 \cos \left(4 \theta _3\right) \cos ^2\left(2 \theta _5\right)-\cos \left(4 \theta _5\right)+3\right),
    \nonumber\\
    \gamma_\pm &= \frac{1}{2} \sqrt{4\pm \sqrt{\cos \left(4 \theta _5\right)+\cos \left(4 \theta _3\right) \left(\cos \left(4 \theta _5\right)+1\right)+13}}.
\end{align}

\subsection{Quantum generative adversarial network (QGAN) ansatz}

The output state of the quantum generative adversarial network (QGAN) depicted by Fig. \ref{fig:circuit_diagrams_v2}(b) is
\begin{align}
  \ket{\psi}_{QGAN} &=   e^{-\frac{1}{2} i \left(\theta _3+\theta _4+\theta _5\right)} \cos\!
\left(\frac{\theta _1}{2}\right) \cos\! \left(\frac{\theta _2}{2}\right) \ket{00}
-i e^{-\frac{1}{2} i \left(\theta _3-\theta _4-\theta _5\right)} \sin \left(\frac{\theta
   _2}{2}\right) \cos \left(\frac{\theta _1}{2}\right)
\ket{01} \nonumber\\
&\quad -i e^{\frac{1}{2} i \left(\theta _3-\theta _4+\theta _5\right)} \sin \left(\frac{\theta
   _1}{2}\right) \cos \left(\frac{\theta _2}{2}\right) \ket{10} 
-e^{\frac{1}{2} i \left(\theta _3+\theta _4-\theta _5\right)} \sin \left(\frac{\theta
   _1}{2}\right) \sin \left(\frac{\theta _2}{2}\right)
   \ket{11}
\end{align}

The angle parameters of the $S^4$ Hopf base $(\theta_A,\phi_A)$ and $(\chi,\xi)$ and the equivalent Cartesian coordinates $(x_0,x_1,x_2,x_3,x_4)$ are calculated to be
\begin{align}
    x_0 &= \cos \left(\theta _1\right)\\
    x_1 &= \sin \left(\theta _1\right) \left[\sin \left(\theta _3\right) \cos
\left(\theta _5\right)+\sin \left(\theta _5\right) \cos \left(\theta_2\right) \cos \left(\theta _3\right)\right] \\
x_2 &= -\sin \left(\theta _1\right) \sin \left(\theta _2\right) \sin \left(\theta _5\right) \\
x_3 &= 0 \\
x_4 &= \sin \left(\theta _1\right) \left[\sin \left(\theta _3\right) \sin \left(\theta _5\right) \cos \left(\theta _2\right)-\cos \left(\theta _3\right) \cos \left(\theta _5\right)\right]
\end{align}
and
\begin{align}
    \theta_A &= \theta_1 \\
    \phi_A &= \arccos\left[\sin \left(\theta _3\right) \cos \left(\theta _5\right)+\sin \left(\theta _5\right) \cos \left(\theta _2\right) \cos \left(\theta _3\right)\right] \\
    \chi &= \arccos\left(\frac{\sin \left(\theta _3\right) \sin \left(\theta _5\right) \cos \left(\theta _2\right)-\cos \left(\theta _3\right) \cos \left(\theta _5\right)}{\sqrt{1-\left(\sin \left(\theta _3\right) \cos \left(\theta _5\right)+\sin \left(\theta _5\right) \cos \left(\theta _2\right) \cos \left(\theta _3\right)\right){}^2}}\right)\\
    \xi &= 0
\end{align}

The quaternion $q_\pm=\langle c_\pm|\psi_H\rangle$ is calculated to be
\begin{align}
   \label{qgans_fiber}
    q_\pm = \frac 1{\sqrt 2} \left(\sin \left(\frac{\theta _1}{2}\right) \gamma_\mp+\cos \left(\frac{\theta _1}{2}\right) \gamma_\pm \right)(a_\pm + ib_\pm + jc_\pm+kd_\pm)
\end{align}
where
\begin{align}
    a_\pm &= \cos \left(\frac{\theta _2}{2}\right) \cos \left(\frac{\theta _3+\theta _4+\theta _5}{2}
    \right) 
    \\
    b_\pm &=
    -\cos \left(\frac{\theta _2}{2}\right) \sin \left(\frac{\theta _3+\theta _4+\theta _5}{2}
    \right)
    \\
    c_\pm &=
    -\sin \left(\frac{\theta _2}{2}\right) \sin \left(\frac{\theta _3-\theta _4-\theta _5}{2}
    \right)
    \\
        d_\pm &=
    -\sin \left(\frac{\theta _2}{2}\right) \cos \left(\frac{\theta _3-\theta _4-\theta _5}{2}
    \right)
\end{align}
where
\begin{align}
    \gamma_\pm=
    \frac{1}{2} \sqrt{4\pm \sqrt{14+2\cos \left(2 \theta _1\right)}}.
\end{align}

Where for example $\langle \sigma^z_2\rangle_2 $ is the expectation value of $\sigma^z_2$ after the second layer of layer of the circuit.

\subsection{Sycamore hardware-efficient ansatz (sHEA)}

The output state of the Sycamore hardware-efficient ansatz (sHEA) depicted by Fig. \ref{fig:circuit_diagrams_v2}(d) is
\begin{align}
\ket{\psi}_{sHEA} &=   -i e^{-\frac{1}{2} i \left(\theta _5+\theta _6\right)} \sin \left(\frac{\theta _2}{2}\right) \cos \left(\frac{\theta _1}{2}\right) \ket{00} \nonumber\\
&\quad + e^{-\frac{1}{2} i \left(\theta _5-\theta _6\right)} \left(\cos \left(\frac{\theta _1}{2}\right) \cos \left(\frac{\theta _2}{2}\right) \cos \left(\frac{\theta _3}{2}\right)-i \sin \left(\frac{\theta _1}{2}\right) \sin \left(\frac{\theta _2}{2}\right) \sin \left(\frac{\theta _3}{2}\right)\right)
\ket{01} \nonumber\\
&\quad + e^{\frac{1}{2} i \left(\theta _5-\theta _6\right)} \left(-\sin \left(\frac{\theta _1}{2}\right) \sin \left(\frac{\theta _2}{2}\right) \cos \left(\frac{\theta _3}{2}\right)+i \sin \left(\frac{\theta _3}{2}\right) \cos \left(\frac{\theta _1}{2}\right) \cos \left(\frac{\theta _2}{2}\right)\right) \ket{10} \nonumber\\
&\quad - i e^{-\frac{1}{4} i \left(\theta _4-2 \left(\theta _5+\theta _6\right)\right)} \sin \left(\frac{\theta _1}{2}\right) \cos \left(\frac{\theta _2}{2}\right) 
   \ket{11}
\end{align}

\begin{landscape}
The angle parameters of the $S^4$ Hopf base $(\theta_A,\phi_A)$ and $(\chi,\xi)$ and the equivalent Cartesian coordinates $(x_0,x_1,x_2,x_3,x_4)$ are calculated to be
\begin{align}
    x_0 &= \frac{1}{2} \left(\cos \left(\theta _2\right) \left(\cos \left(\theta _3\right)-1\right)+\cos \left(\theta _1\right) \left(\cos \left(\theta _3\right)+1\right)\right)\\
    x_1 &= \sin \left(\theta _2\right) \sin \left(\frac{\theta _3}{2}\right) \left(\sin ^2\left(\frac{\theta _1}{2}\right) \cos \left(\frac{\theta _4}{4}-\theta _5\right)-\cos ^2\left(\frac{\theta _1}{2}\right) \cos \left(\theta _5\right)\right) \nonumber\\
    &\quad + \sin \left(\theta _1\right) \cos \left(\frac{\theta _3}{2}\right) \left(\sin ^2\left(\frac{\theta _2}{2}\right) \sin \left(\theta _5\right)-\sin \left(\frac{\theta _4}{4}-\theta _5\right) \cos ^2\left(\frac{\theta _2}{2}\right)\right)\\
    x_2 &= \frac{1}{2} \left(\sin \left(\theta _3\right)-\sin \left(\theta _1\right) \sin \left(\theta _2\right) \sin \left(\frac{\theta _4}{4}\right)+\sin \left(\theta _3\right) \cos \left(\theta _1\right) \cos \left(\theta _2\right)\right) \\
    x_3 &= \frac{1}{2} \sin \left(\theta _1\right) \sin \left(\theta _2\right) \left(\cos \left(\theta _3\right)-\cos \left(\frac{\theta _4}{4}\right)\right)\\
    x_4 &= \frac{1}{2} \bigg(\sin (\theta _1) \cos \bigg(\frac{\theta _3}{2}\bigg) \bigg(-\bigg(-\cos (\theta _2)+\bigg(\cos (\theta _2)+1\bigg) \cos \bigg(\frac{\theta _4}{4}\bigg)+1\bigg) \cos (\theta _5) \nonumber\\ 
    &\quad -\sin \bigg(\frac{\theta _4}{4}\bigg) \sin (\theta _5) \bigg(\cos (\theta _2)+1\bigg)\bigg)-\sin(\theta _2) \sin \bigg(\frac{\theta _3}{2}\bigg) \bigg(2 \sin \bigg(\frac{\theta _4}{4}\bigg) \sin ^2\bigg(\frac{\theta _1}{2}\bigg) \cos (\theta _5)\nonumber\\ &\quad +\sin (\theta _5) \bigg(\cos (\theta _1) +\big(\cos (\theta _1)-1\big) \cos \bigg(\frac{\theta _4}{4}\bigg)+1\bigg)\bigg)\bigg)
\end{align}
and
\begin{align}
    \theta_A &= \arccos \left[ \frac{1}{2} \left(\cos \left(\theta _2\right) \left(\cos \left(\theta _3\right)-1\right)+\cos \left(\theta _1\right) \left(\cos \left(\theta _3\right)+1\right)\right) \right] \\
    \phi_A &= \arccos \left[ 
    \frac{
        \splitfrac{
        \sin \left(\theta _2\right) \sin \left(\frac{\theta _3}{2}\right) \left(\sin ^2\left(\frac{\theta _1}{2}\right) \cos \left(\frac{\theta _4}{4}-\theta _5\right)-\cos ^2\left(\frac{\theta _1}{2}\right) \cos \left(\theta _5\right)\right)
        }{
        +\sin \left(\theta _1\right) \cos \left(\frac{\theta _3}{2}\right) \left(\sin ^2\left(\frac{\theta _2}{2}\right) \sin \left(\theta _5\right)-\sin \left(\frac{\theta _4}{4}-\theta _5\right) \cos ^2\left(\frac{\theta _2}{2}\right)\right)
        }
    }{
    \sqrt{1-\frac{1}{4} \left(\cos \left(\theta _2\right) \left(\cos \left(\theta _3\right)-1\right)+\cos \left(\theta _1\right) \left(\cos \left(\theta _3\right)+1\right)\right){}^2}
    } \right]\\
    \chi &= \arccos \left[ 
    \frac{
        \splitfrac{
            \sin \left(\theta _1\right) \cos \left(\frac{\theta _3}{2}\right) \left(-\left(-\cos \left(\theta _2\right)+\left(\cos \left(\theta _2\right)+1\right) \cos \left(\frac{\theta _4}{4}\right)+1\right) \cos \left(\theta _5\right)
            -\sin \left(\frac{\theta _4}{4}\right) \sin \left(\theta _5\right) \left(\cos \left(\theta _2\right)+1\right)\right)
        }{
            -\sin \left(\theta _2\right) \sin \left(\frac{\theta _3}{2}\right) \left(2 \sin \left(\frac{\theta _4}{4}\right) \sin ^2\left(\frac{\theta _1}{2}\right) \cos \left(\theta _5\right)
            +\sin \left(\theta _5\right) \left(\cos \left(\theta _1\right)+\left(\cos \left(\theta _1\right)-1\right) \cos \left(\frac{\theta _4}{4}\right)+1\right)\right)
        }
    }{
    \splitfrac{
    2 \left( 
    \left( 1-\frac{1}{4} \left(\cos \left(\theta _2\right) \left(\cos \left(\theta _3\right)-1\right)+\cos \left(\theta _1\right) \left(\cos \left(\theta _3\right)+1\right)\right){}^2\right)
    \right.
    }{
    \left.
    \times \left(1-\frac{\left(\sin \left(\theta _2\right) \sin \left(\frac{\theta _3}{2}\right) \left(\sin ^2\left(\frac{\theta _1}{2}\right) \cos \left(\frac{\theta _4}{4}-\theta _5\right)-\cos ^2\left(\frac{\theta _1}{2}\right) \cos \left(\theta _5\right)\right)+\sin \left(\theta _1\right) \cos \left(\frac{\theta _3}{2}\right) \left(\sin ^2\left(\frac{\theta _2}{2}\right) \sin \left(\theta _5\right)-\sin \left(\frac{\theta _4}{4}-\theta _5\right) \cos ^2\left(\frac{\theta _2}{2}\right)\right)\right){}^2}{1-\frac{1}{4} \left(\cos \left(\theta _2\right) \left(\cos \left(\theta _3\right)-1\right)+\cos \left(\theta _1\right) \left(\cos \left(\theta _3\right)+1\right)\right){}^2}\right)
    \right)^{-1}
    }
    } \right] \\
    \xi &= \arctan \left[ \frac{\sin \left(\theta _1\right) \sin \left(\theta _2\right) \left(\cos \left(\theta _3\right)-\cos \left(\frac{\theta _4}{4}\right)\right)}{\sin \left(\theta _3\right)-\sin \left(\theta _1\right) \sin \left(\theta _2\right) \sin \left(\frac{\theta _4}{4}\right)+\sin \left(\theta _3\right) \cos \left(\theta _1\right) \cos \left(\theta _2\right)} \right]
\end{align}

The quaternion $q_\pm=\langle c_\pm|\psi_H\rangle$ is calculated to be
\begin{align}
   \label{shea_fiber}
    q_\pm =
    m_\pm + im_\pm + jr_\pm+ks_\pm
\end{align}
where
\begin{align}
    m_\pm &= \frac{\sin \left(\frac{\theta _2}{2}\right) \sin \left(\frac{1}{2} \left(\theta _5+\theta _6\right)\right) \cos \left(\frac{\theta _1}{2}\right) \left(\frac{2 \left(\cos \left(\theta _2\right) \left(\cos \left(\theta _3\right)-1\right)+\cos \left(\theta _1\right) \left(\cos \left(\theta _3\right)+1\right)-2\right) \gamma_{\mp}}{\sqrt{-4 \cos \left(2 \theta _1\right) \cos ^4\left(\frac{\theta _3}{2}\right)-\cos \left(2 \theta _3\right)-4 \sin ^4\left(\frac{\theta _3}{2}\right) \cos \left(2 \theta _2\right)+4 \sin ^2\left(\theta _3\right) \cos \left(\theta _1\right) \cos \left(\theta _2\right)+5}}-\sqrt{2} \gamma_{\pm} \right)}{\sqrt{2}}
    \\
    r_\pm &= 
    \frac{
    \left(\cos \left(\frac{1}{2} \left(\theta _1+\theta _2-\theta _3+\theta _5-\theta _6\right)\right)+\cos \left(\frac{1}{2} \left(\theta _1-\theta _2+\theta _3+\theta _5-\theta _6\right)\right)+2 \cos \left(\frac{1}{2} \left(\theta _2+\theta _3\right)\right) \cos \left(\frac{1}{2} \left(\theta _1-\theta _5+\theta _6\right)\right)\right) 
    \left(
    - B_\mp + 
    \sqrt{2} \gamma_\pm 
    C
    \right)
    }{
    A
    }
    \\
    s_\pm &= 
    \frac{
    \left(\sin \left(\frac{1}{2} \left(\theta _1+\theta _2-\theta _3+\theta _5-\theta _6\right)\right)+\sin \left(\frac{1}{2} \left(\theta _1-\theta _2+\theta _3+\theta _5-\theta _6\right)\right)-2 \sin \left(\frac{1}{2} \left(\theta _1-\theta _5+\theta _6\right)\right) \cos \left(\frac{1}{2} \left(\theta _2+\theta _3\right)\right)\right)
    \left(
    B_\mp - 
    \sqrt{2} \gamma_\pm 
    C
    \right)
    }{
    A
    }
\end{align}

where
\begin{align}
A &= 4 \sqrt{2} \sqrt{-4 \cos \left(2 \theta _1\right) \cos ^4\left(\frac{\theta _3}{2}\right)-\cos \left(2 \theta _3\right)-4 \sin ^4\left(\frac{\theta _3}{2}\right) \cos \left(2 \theta _2\right)+4 \sin ^2\left(\theta _3\right) \cos \left(\theta _1\right) \cos \left(\theta _2\right)+5},
\end{align}
\begin{align}
B_\mp &= 2 \left(\cos \left(\theta _2\right) \left(\cos \left(\theta _3\right)-1\right)+\cos \left(\theta _1\right) \left(\cos \left(\theta _3\right)+1\right)-2\right) \gamma_{\mp},
\end{align}
\begin{align}
C &= \sqrt{-4 \cos \left(2 \theta _1\right) \cos ^4\left(\frac{\theta _3}{2}\right)-\cos \left(2 \theta _3\right)-4 \sin ^4\left(\frac{\theta _3}{2}\right) \cos \left(2 \theta _2\right)+4 \sin ^2\left(\theta _3\right) \cos \left(\theta _1\right) \cos \left(\theta _2\right)+5},
\end{align}
and
\begin{align}
    \gamma_\pm=
    \frac{1}{2} \sqrt{\frac{\pm \sqrt{4 \cos \left(2 \theta _1\right) \cos ^4\left(\frac{\theta _3}{2}\right)+\cos \left(2 \theta _3\right)+4 \sin ^4\left(\frac{\theta _3}{2}\right) \cos \left(2 \theta _2\right)-4 \sin ^2\left(\theta _3\right) \cos \left(\theta _1\right) \cos \left(\theta _2\right)+27}}{\sqrt{2}}+4}.
\end{align}

\end{landscape}

\section{Geometric origins of the quantum natural gradient}

Considering the geometric point of view, how might one arrive at the concept of the quantum natural gradient? 
The idea is to simply find out how the gradient operation would change in different curved geometries. 
For the quantum mechanics, we would need to consider the geometry of $\mathbb{P}(H)$. 

Using the correspondence between a vector space $V$ and its dual $V^*$ for finite dimensional spaces, we can pair the co-vector $df$ with the vector $\nabla f$ i.e 
 \begin{equation}
    \label{gradientdefinition}
     df(w)= \langle \nabla f, w \rangle
 \end{equation}
 where  $\langle \cdot , \cdot \rangle$ is a chosen bilinear form. Now consider the following calculation:
 \begin{equation*}
     \langle v, w \rangle = \sum_{ij} v^{i} \langle e_i, e_j \rangle w^j = \nu (w),
 \end{equation*}
 where $v, w \in V$ and $\nu \in V^*$ and $ \{ e_i\}$ are basis for $V$. We can expand $\nu$ in a basis $\{\sigma^i\}$ for $V^*$ so that $\nu = \sum_i v_i \sigma^i $ with $v_i =\nu(e_i)= \langle v, e_j \rangle $. From the fact $v^i = \sum_{j}g^{ij}v_j = \sum_i g^{ij}\langle v, e_j \rangle$, we have
 \begin{equation}
     \label{transformation_rule}
     v = \sum_{j} v^i e_i = \sum_i \left( \sum_j g^{ij} \langle v, e_j \rangle \right) e_i .
 \end{equation}
 
 By the discussion above, we have that 
 \begin{equation}
     \label{gradient_coordinate_definition}
     \nabla f = \sum_i (\nabla f)^i e_i = \sum_i \left( \sum_j g^{ij} \langle \nabla f, e_i \rangle \right) e_i.
 \end{equation}
 
 Considering (\ref{gradientdefinition}) we have that $df(w)= \langle \nabla f, w \rangle = w(f) = \sum_{j} w^j \frac{\partial f}{\partial x^j} $ so that combining knowledge from (\ref{transformation_rule}) and (\ref{gradient_coordinate_definition})  we have $\langle \nabla , e_j \rangle = e_j(f)=\frac{\partial f}{\partial x^j} $ and thus arrive at 
 \begin{equation}
    (\nabla f)^i = \sum_j g^{ij} \frac{\partial f}{\partial x^j}.
 \end{equation}

Now in Euclidean geometry $df$ and $\nabla f$ have the same co-ordinates but in general they do not. For $\mathbb{P}(H)$ we pick the metric to be Fubini-Study metric and thus arrive at the following update rule for the parameters:
 
\begin{equation} 
\label{natural_gradient_descent}
    \boldsymbol\theta^{t+1} = \boldsymbol\theta^{t} - \eta g^{-1} \frac{\partial \mathcal{L}(\boldsymbol\theta)}{\partial \boldsymbol\theta}.
\end{equation}

\section{Further insights into VQE performance through scalar curvature: unentangled ground state}\label{app:50_runs_product}

\begin{figure}[H]
\centering
\includegraphics[scale=0.32]{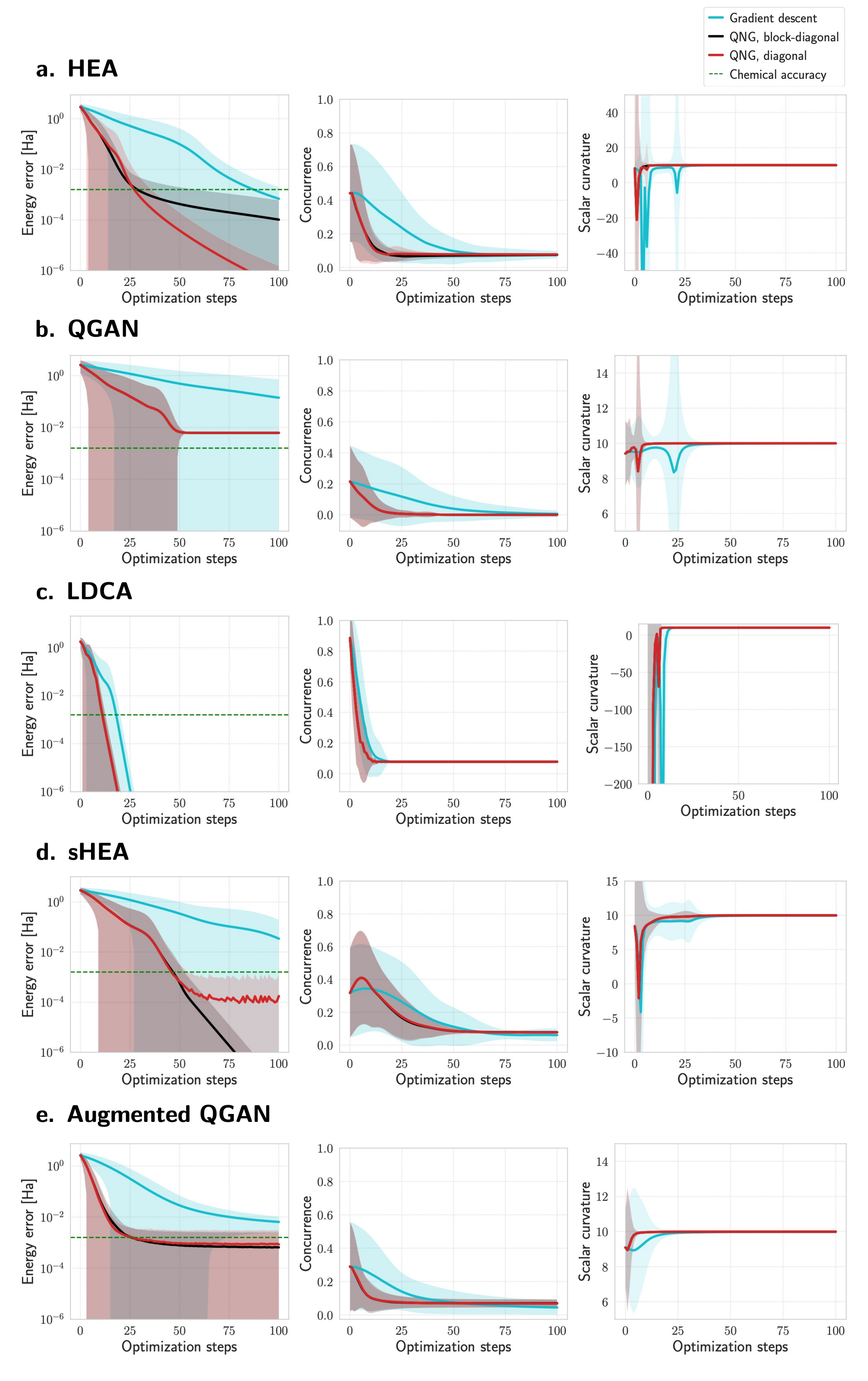}
\caption{
VQE results over 50 trials using the four two-qubit ansatze for molecular hydrogen where $R_{\text{H-H}} = 0.2 \ \angstrom$.
Each ansatz is optimized using gradient descent and quantum natural gradient (QNG) using the block-diagonal and diagonal approximations to the metric tensor.
Lines in each plot indicate the average quantity, and shaded regions indicate quantities within one standard deviation.
}
\label{fig:h2_separable_ground_state_data_50_runs}
\end{figure}

In this section, we consider another regime of the same VQE problem from Sec. \ref{sec:numerical_experiments}.
At $r_{H-H} = 0.2 \angstrom$, the ground state wave function is a product state:
\begin{equation}
    \ket{\psi_g} \approx \beta \ket{10},
\end{equation}
with $|\beta|^2 \approx 1$.
We repeat the simulation procedure, running 50 independent VQE optimizations for each of the four two-qubit circuits, as shown in Fig.~\ref{fig:h2_separable_ground_state_data_50_runs}.
We observe similar results for $r_{H-H} = 0.2 \angstrom$ as those for $r_{H-H} = 3.19 \angstrom$; using LDCA, the optimizations are both precise and accurate.
While initial concurrence values start at high values, they rapidly decrease to near 0. This corresponds to starting in a negative curvature region but quickly moving up to a hill of positive curvature.
QGAN, in its original structure, is again insufficient for converging to the ground state energy. However, with added local rotations (Fig.~\ref{fig:h2_separable_ground_state_data_50_runs}e), the optimizations reach sufficient accuracy. 

\newpage
\bibliographystyle{unsrt}
\bibliography{references}

\end{document}